\documentclass[prd,twocolumn,showpacs,preprintnumbers,amsmath,amssymb,nofootinbib]{revtex4}

\usepackage{graphicx}
\usepackage{dcolumn}
\usepackage{bm}
\usepackage{color}
\usepackage{units}
\usepackage{amsmath}

\usepackage[caption=false]{subfig}
		
\usepackage[colorlinks=true,
	linkcolor=black,
	citecolor=black,
	urlcolor=black,
	filecolor=black
]{hyperref}

\newcommand{\eq}[1]{Eq.~(\ref{#1})}

\begin{document}

\preprint{\vbox{\hbox{PSI-PR-16-04}}} 
\preprint{\vbox{\hbox{FERMILAB-PUB-16-131-T}}}

\title{\boldmath A $L_\mu - L_\tau$ theory of Higgs flavor violation and  $(g-2)_\mu$}

\author{Wolfgang Altmannshofer}
\email{altmanwg@ucmail.uc.edu}
\affiliation{Department of Physics, University of Cincinnati, Cincinnati, Ohio 45221,USA}

\author{Marcela Carena}
\email{carena@fnal.gov}
\affiliation{Fermilab, P.O. Box 500, Batavia, IL 60510, USA}
\affiliation{Enrico Fermi Institute, University of Chicago, Chicago, IL 60637, USA}
\affiliation{Kavli Institute for Cosmological Physics,University of Chicago, Chicago, IL 60637, USA}

\author{Andreas Crivellin}
\email{andreas.crivellin@cern.ch}
\affiliation{Paul Scherrer Institut, CH--5232 Villigen PSI, Switzerland}

\begin{abstract}
Several experiments reported hints for the violation of lepton flavor or lepton flavor universality in processes involving muons. Most prominently, there is the hint for a non-zero rate of the flavor violating Higgs decay $h \to \tau\mu$ at the LHC, as well as the hint for lepton flavor universality violation in rare $B$ meson decays at LHCb. In addition, also the long standing discrepancy in the anomalous magnetic moment of the muon motivates new physics connected to muons. A symmetry which violates lepton flavor universality, is $L_\mu-L_\tau$: the difference of muon-number and tau-number. We show that adding vector-like fermions to a $L_\mu-L_\tau$ theory generates naturally an effect in the anomalous magnetic moment of the muon and $h\to\tau\mu$, while effects in other $\tau \to \mu$ transitions are systematically suppressed by symmetry arguments. We find that if $L_\mu-L_\tau$ is gauged it is possible to also accommodate the discrepant 
$b\to s\mu\mu$ data while predicting a $\tau\to3\mu$ and a modified $h\to\mu\mu$ rate within reach 
of upcoming experiments.
\end{abstract}

\pacs{12.60.Cn, 12.60.Fr, 13.35.Dx, 13.40.Em, 14.80.Ec}

\maketitle

{\em Introduction.}
%
In the Standard Model (SM) of particle physics lepton flavor universality (LFU) is only violated by Higgs Yukawa interactions and lepton flavor is even conserved (neglecting extremely tiny neutrino mass effects). Any observation of the lepton flavor violation (LFV) would be a clear sign for physics beyond the SM and also evidence for LFU violation (LFUV), beyond the Yukawa interactions, would indicate new physics (NP).

A prominant LFV process at the LHC is the flavor violating Higgs decay $h\to\tau\mu$~\cite{Davidson:2010xv,Blankenburg:2012ex,Harnik:2012pb}. Interestingly, the CMS collaboration found a mild excess of $2.4\sigma$ in the search for $h\to\tau\mu$~\cite{Khachatryan:2015kon} corresponding to
\begin{align}
{\rm BR} (h\to\tau\mu) = \left( 0.84_{-0.37}^{+0.39} \right)\% \,.
\label{h0taumuExp}
\end{align}
This excess is consistent with the less sensitive ATLAS analyses~\cite{Aad:2016blu}, that find observed limits weaker than the expected ones. However, searches for other LFV processes such as $\mu\to e\gamma$, $\tau\to \mu \gamma$, or $\tau\to 3\mu$ have so far been unsuccessful and put stringent limits on the presence of new sources of LFV. Possible explanations of the $h\to\tau\mu$ signal that are not in conflict with the null results from other searches for charged LFV generically contain an extended Higgs sector, i.e. new sources of electroweak symmetry breaking (see for example Refs.~\cite{Campos:2014zaa,Sierra:2014nqa,Heeck:2014qea,Crivellin:2015mga,Dorsner:2015mja,Omura:2015nja,Crivellin:2015lwa,Chiang:2015cba,Altmannshofer:2015esa,Aloni:2015wvn,Alvarado:2016par}).

Hints for the violation of $\mu-e$ universality in rare $B$ meson decays have been reported by LHCb.\footnote{There are also hints for LFUV in semi-tauonic $B$ decays~\cite{Amhis:2014hma}. Explanations in terms of BSM physics require new charged current interactions and will not be discussed here.} In particular, the LFU ratio $R_K$
\begin{equation}
R_K=\frac{\text{BR}(B\to K \mu^+\mu^-)}{\text{BR}(B\to K e^+e^-)}= 0.745^{+0.090}_{-0.074}\pm 0.036\,,
\end{equation}
has been measured for a dilepton invariant mass in the range $1\,{\rm GeV^2}<q^2<6\,{\rm GeV^2}$ by LHCb~\cite{Aaij:2014ora}. The measured value of $R_K$ disagrees with the theoretically clean SM prediction $R_K^{\rm SM}=1.0003 \pm 0.0001$~\cite{Bobeth:2007dw} by $2.6\,\sigma$. Assuming that NP affects only the muon mode but not the electron mode, the anomaly in $R_K$ is compatible with other anomalies in rare $b \to s \mu\mu$ decays and a combined fit prefers NP to the SM by $4-5\,\sigma$~\cite{Altmannshofer:2014rta,Altmannshofer:2015sma,Descotes-Genon:2015uva,Hurth:2016fbr} depending on assumptions made for the hadronic uncertainties~\cite{Lyon:2014hpa,Descotes-Genon:2014uoa,Altmannshofer:2014rta,Jager:2014rwa}. This situation is naturally realized in models with gauged muon-number minus tau-number ($L_\mu-L_\tau$)~\cite{Altmannshofer:2014cfa,Crivellin:2015mga,Crivellin:2015lwa,Altmannshofer:2015mqa}, where a $Z^\prime$ gauge boson gives tree level contributions to $b \to s \mu\mu$ transitions but leaves 
the electron channel unaffected\footnote{New physics explanations can also be obtained in other $Z^\prime$ models~\cite{Niehoff:2015bfa,Sierra:2015fma,
Celis:2015ara,Belanger:2015nma,Falkowski:2015zwa,Carmona:2015ena,Buras:2016dxz}. Alternative explanations are models with leptoquarks~\cite{Gripaios:2014tna,Becirevic:2015asa,Varzielas:2015iva,Alonso:2015sja,Calibbi:2015kma,Bauer:2015knc,Dorsner:2016wpm}}. 

The anomalous magnetic moment (AMM) of the muon $a_\mu \equiv (g-2)_\mu/2$, provides another motivation for NP connected to muons. The experimental value of $a_\mu$ is completely dominated by the Brookhaven experiment E821~\cite{Bennett2006} and is given by~\cite{Olive2014} $a_\mu^\mathrm{exp} = (116\,592\,091\pm54\pm33) \times 10^{-11}$, where the first error is statistical and the second systematic. The SM prediction is~\cite{Aoyama2012,Czarnecki1995,Czarnecki1996,Gnendiger2013,Davier2011,Hagiwara2011,Kurz2014,Jegerlehner2009,Colangelo2014} $a_\mu^\mathrm{SM} = (116\,591\,855\pm59) \times 10^{-11}$, where almost the entire uncertainty is due to hadronic effects. This amounts to a discrepancy between the SM and experimental values of
\begin{equation} \label{eq:g-2}
 \Delta a_\mu = a_\mu^\mathrm{exp}-a_\mu^\mathrm{SM} = (236\pm 87)\times 10^{-11}\, ,
\end{equation}
i.e.~a $2.7\sigma$ deviation\footnote{Less conservative estimates lead to discrepancies up to $3.6\,\sigma$}. Possible NP explanations besides supersymmetry (see for example Ref.~\cite{Stockinger:2006zn} for a review) include leptoquarks~\cite{Chakraverty:2001yg,Cheung:2001ip}, additional fermions \cite{Freitas:2014pua}, new scalar contributions in two-Higgs-doublet models (2HDM)~\cite{Iltan:2001nk,Omura:2015nja}, also within the lepton-specific 2HDM~\cite{Broggio:2014mna,Wang:2014sda,Abe:2015oca,Crivellin:2015hha}, and very light $Z^\prime$ bosons~\cite{Langacker:2008yv,Baek:2001kca,Ma:2001md,Gninenko:2001hx,Pospelov:2008zw,Heeck:2011wj,Harigaya:2013twa,Altmannshofer:2014pba}, in particular the $Z^\prime$ gauge boson related to gauging $L_\mu - L_\tau$. 

The abelian $L_\mu-L_\tau$ symmetry is interesting in general: not only is it an anomaly-free global symmetry within the SM~\cite{He:1990pn,Foot:1990mn,He:1991qd}, it also leads to a good zeroth-order approximation for neutrino mixing with a quasi-degenerate $\nu_\mu$, $\nu_\tau$ mass spectrum, predicting a maximal atmospheric and vanishing reactor neutrino mixing angle~\cite{Binetruy:1996cs,Bell:2000vh,Choubey:2004hn}. Breaking of $L_\mu-L_\tau$ is mandatory for a realistic neutrino sector, and such a breaking can also induce charged LFV processes, such as $\tau\to3\mu$~\cite{Dutta:1994dx,Heeck:2011wj} and $h\to \mu \tau$~\cite{Heeck:2014qea}.

In this Letter we extend the basic $L_\mu-L_\tau$ model by including vector-like leptons which are neutral under $L_\mu-L_\tau$. We find that in this framework the $h \to \tau\mu$ signal can be naturally explained without violating bounds from $\tau \to \mu\gamma$. At the same time one can account for the AMM of the muon and for LFUV in rare $B$ decays.

\bigskip
{\em The Model.}
%
We consider a gauged $L_\mu-L_\tau$ model supplemented with one generation of heavy vector-like leptons. The $L_\mu-L_\tau$ symmetry amounts to assigning charge $+1$ to muons (and muon neutrinos), charge $-1$ to taus (and tau neutrinos) while keeping electrons (and electron neutrinos) uncharged. We choose the vector-like leptons to be neutral under $L_\mu-L_\tau$. If one aims at an explanation of the $b\to s\mu\mu$ anomalies, one can in addition introduce vector-like quarks with appropriate $L_\mu-L_\tau$ charges, as shown in Ref.~\cite{Altmannshofer:2014cfa}.

The $L_\mu-L_\tau$ symmetry is broken spontaneously in a scalar sector. Besides the SM Higgs doublet $H$, it contains a SM singlet scalar $\phi_1$ that carries $L_\mu-L_\tau$ charge $-1$ and a second SM singlet scalar $\phi_2$ also charged under $L_\mu - L_\tau$. For reasons which will become clear later, we assume that the $Z^\prime$ mass originates to a good approximation from only one of the scalars, $\phi_2$. Assuming negligible mixing among the scalars and no couplings of $\phi_2$ with the vector-like leptons, muons and taus (which can be easily achieved by an appropriate charge assignment), the only role of $\phi_2$ is to provide the $Z^\prime$ mass, $m_{Z^\prime}$, which we will therefore treat as independent parameter. The only scalar that is relevant for the charged lepton phenomenology is then $\phi_1$ for which we drop the subscript in the following: $\phi_1\to \phi$.

The vector-like leptons $L$ and $E$ (with the quantum numbers of the SM lepton doublets and the lepton singlets, respectively) have vector-like mass terms
\begin{equation}
\mathcal L_M = - M_L \bar L_L L_R  - M_E \bar E_L E_R ~+~ \text{h.c.} \,.
\end{equation}
The Yukawa couplings of the vector-like leptons to the Higgs doublet are given by
\begin{equation}
\mathcal L_Y =  - Y_{LE} \bar L_L H E_R  - Y_{EL} \bar L_R H E_L  ~+~ \text{h.c.} \,.
\end{equation}
The vector-like leptons can also couple to muons, taus and the SM singlet $\phi$
\begin{eqnarray}
\mathcal L_\lambda &=& -\lambda_{\mu L} \bar\mu_L L_R \phi^* - \lambda_{\tau L} \bar\tau_L L_R \phi \nonumber\\
&& - \lambda_{\mu E} \bar\mu_R E_L \phi^* - \lambda_{\tau E} \bar\tau_R E_L \phi ~+~ \text{h.c.} \,.
\end{eqnarray}
The masses $M_L$ and $M_E$, the Yukawa couplings $Y_{LE}$ and $Y_{EL}$, as well as the couplings $\lambda_i$ can in principle be complex. For simplicity, we will assume them to be real in the following\footnote{Note that the couplings $\lambda_i$ are absent for electrons, as electrons are not charged under $L_\mu - L_\tau$.
However, for electrons one could consider mixing with the vector-like leptons originating from Yukawa couplings to the SM Higgs doublet, or from vector-like mass terms. We assume such terms to be absent, motivated by the tiny electron mass. This could for example be enforced by a global flavor symmetry under which electrons are charged.}.

The Higgs doublet $H$ and the scalar $\phi$ acquire vacuum expectation values $v \approx 174$~GeV and $v_\phi$. In the broken phase we parameterize the neutral components of the Higgs and the scalar as $H^0 = v + ( h + i G^0)/\sqrt{2}$ and $\phi = v_\phi + ( \varphi + i a)/\sqrt{2}$\,, where $\varphi$ ($a$) is the CP-even (CP-odd) component of the scalar, $h$ will become the main component of the 125~GeV Higgs and $G^0$ provides the longitudinal component of the SM $Z$ boson. If $\phi$ was the only source of $L_\mu - L_\tau$ breaking, $a$ would become the longitudinal component of the $Z^\prime$. However, provided that $\phi$ gives only a subdominant contribution of the $Z^\prime$ mass, $a$ remains a physical degree of freedom and is a mass eigenstate, to a good approximation. Due to $L_\mu - L_\tau$ breaking, the masses of $\varphi$ and $a$ can be split $m_\varphi^2 - m_a^2 = \mathcal O(v_\phi^2)$.

\begin{figure}[tb]
\centering
\includegraphics[width=0.15\textwidth]{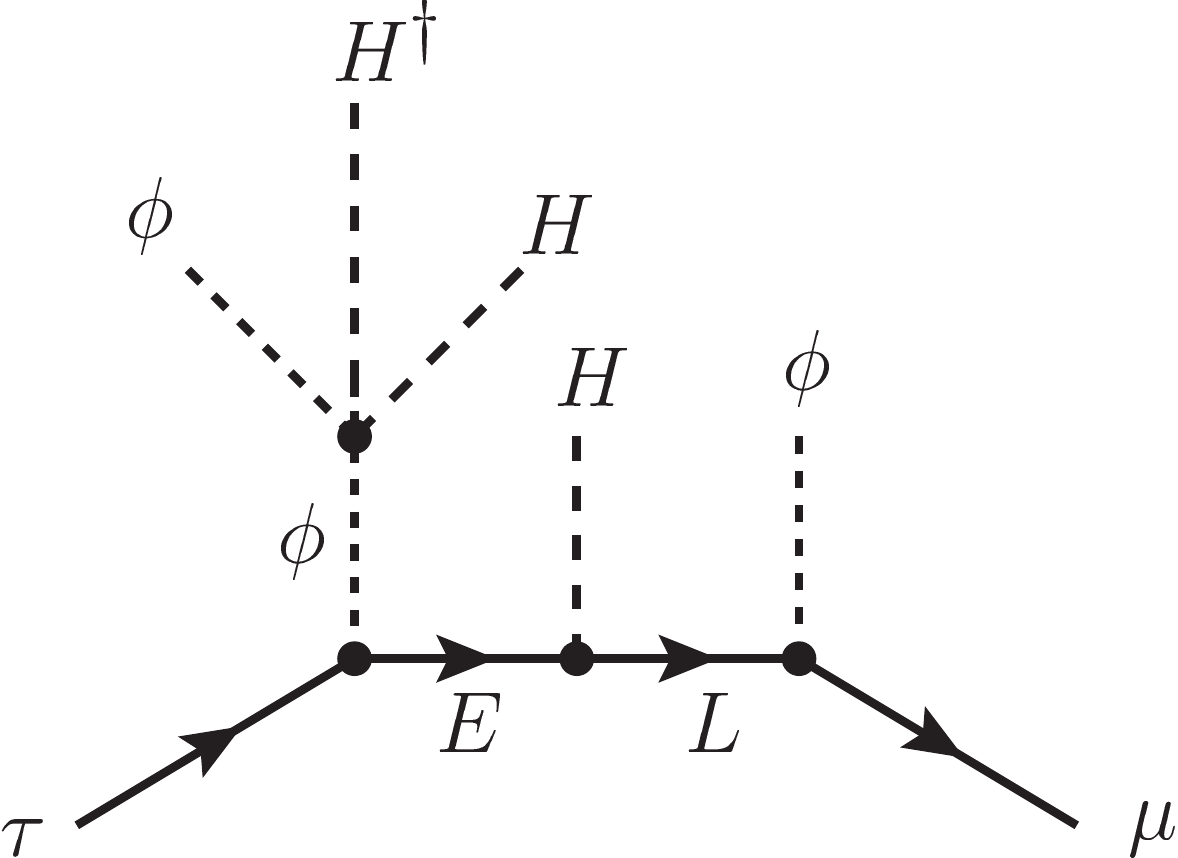} ~
\includegraphics[width=0.15\textwidth]{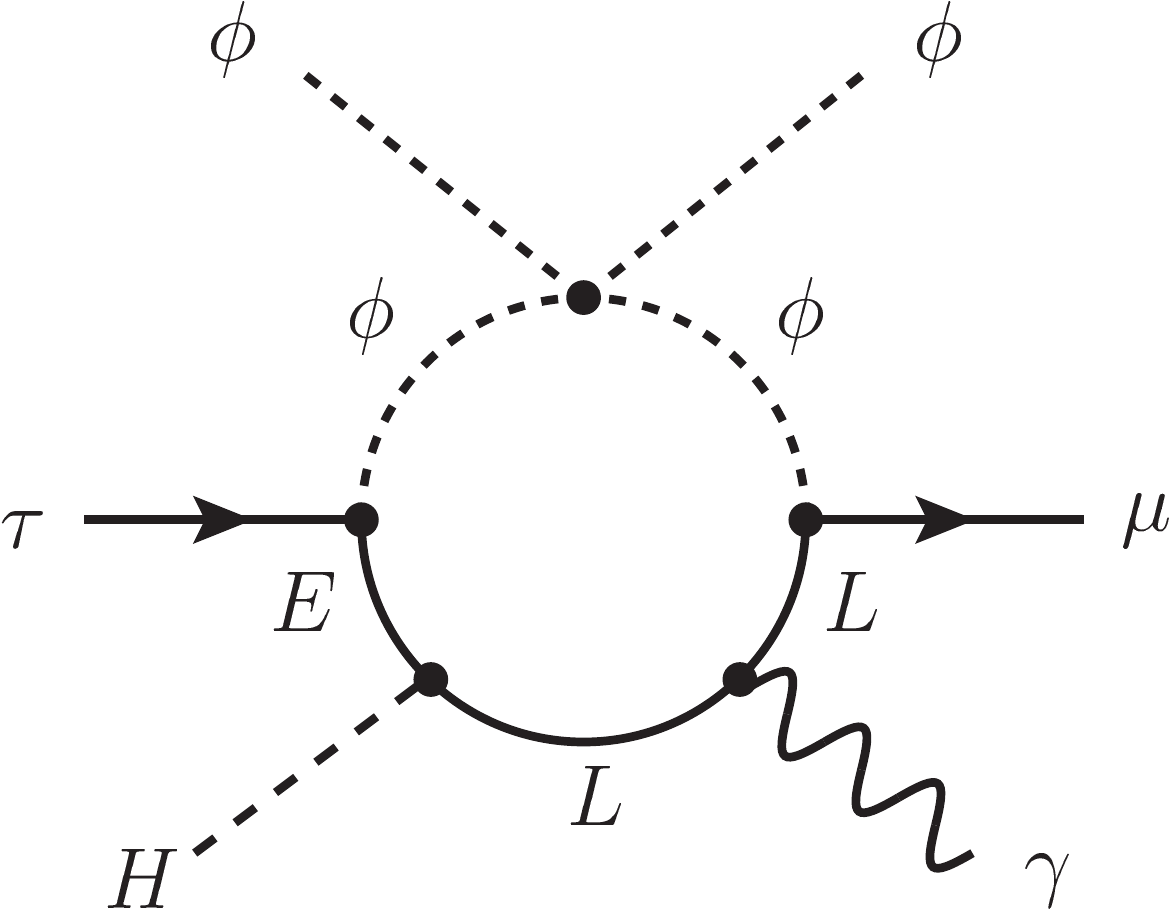} ~
\includegraphics[width=0.15\textwidth]{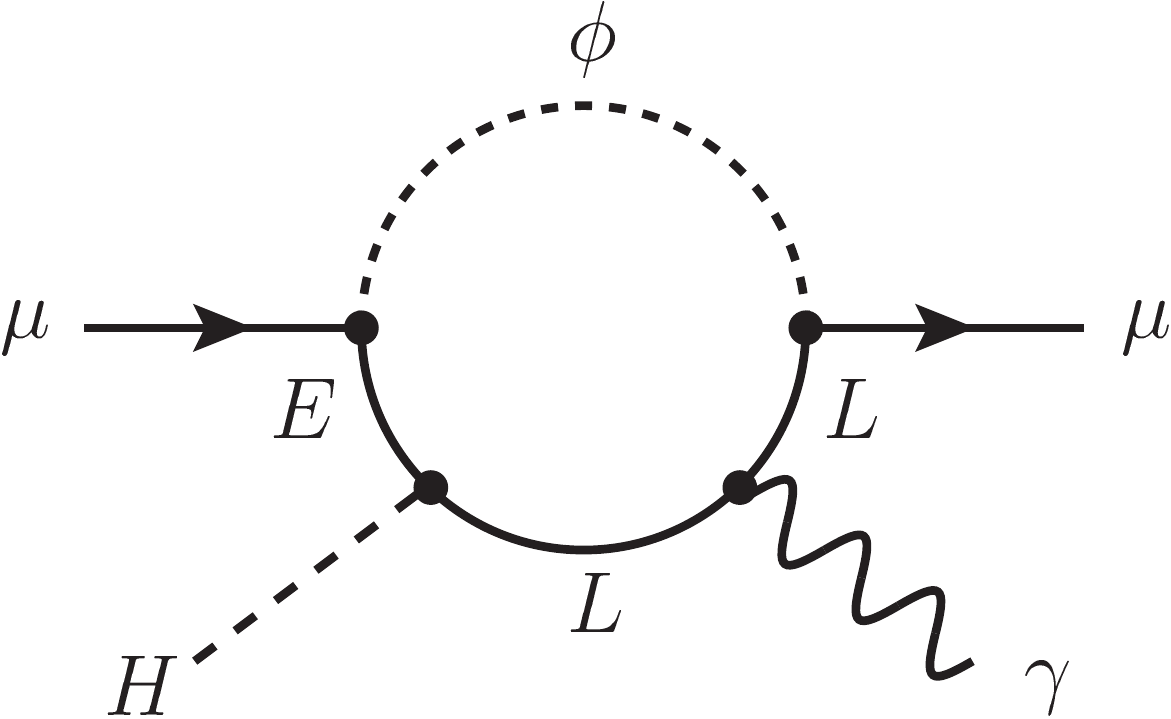}
\caption{Example diagrams giving rise to $h\to\tau\mu$ (left), $\tau\to\mu\gamma$ (center) and $(g-2)_\mu$ (right) at leading order in $v$ and $v_\phi$ after symetry breaking, i.e. after the external $H$ and $\phi$ fields are replaced by their vevs.}
\label{htaumu_diagram}
\end{figure}

In the broken phase, we obtain a $4\times4$ mass matrix for the vector-like leptons, the muon and the tau
\begin{eqnarray}
\begin{pmatrix} \bar L_L \\ \bar E_L \\ \bar \mu_L \\ \bar \tau_L \end{pmatrix}^\text{T}
\begin{pmatrix}
M_L & v Y_{LE} & 0 & 0 \\
v Y_{EL} & M_E & v_\phi \lambda_{\mu E} & v_\phi \lambda_{\tau E} \\
v_\phi \lambda_{\mu L} & 0 & v Y_\mu & 0 \\
v_\phi \lambda_{\tau L} & 0 & 0 & v Y_\tau 
\end{pmatrix} 
\begin{pmatrix} L_R \\ E_R \\ \mu_R \\ \tau_R \end{pmatrix} \,.
\end{eqnarray}
Rotating to lepton mass eigenstates will affect the couplings of leptons to the $Z$, the $Z^\prime$, the (pseudo) scalar ($a$) $\varphi$ and the Higgs $h$. We parameterize the couplings as 
\begin{eqnarray}
 \mathcal L &\supset& \big( \Gamma_{\psi \psi^\prime}^{VL} (\bar\psi_L \gamma^\alpha \psi_L^\prime) + \Gamma_{\psi \psi^\prime}^{VR} (\bar\psi_R \gamma^\alpha \psi_R^\prime) \big) V_\alpha \nonumber \\
 &&+ \left( \Gamma_{\psi\psi^\prime}^S (\bar \psi_L \psi_R^\prime) + \Gamma_{\psi^\prime\psi}^S (\bar\psi_R \psi_L^\prime) \right) S \,,
\end{eqnarray}
with $V = Z, Z^\prime$, $S = \varphi, a, h$, and $\psi,\psi^\prime = \mu,\tau,L,E$.
After symmetry breaking, $\varphi$ can mix with $h$ via a quartic term $\lambda |\phi|^2 |H|^2 $. While this mixing does not affect the CP-odd components, the CP-even mass matrix has to be diagonalized by
\begin{equation}
 \varphi \to \varphi\cos\alpha - h \sin\alpha ~,~~ h \to h\cos\alpha + \varphi \sin\alpha \,,
\end{equation}
where $\sin\alpha \simeq 2 \lambda v v_\phi / m_\varphi^2$. The mixing has three important effects: (i) it reduces all couplings of $h$ to SM states by a factor of $\cos\alpha$. Given the good agreement of Higgs rate measurements at the LHC~\cite{confnote,Aad:2015gba,Khachatryan:2014jba} with the SM predictions, the size of the mixing is constrained. Using the measured combined signal yield relative to the SM expectation of $\mu = 1.09\pm 0.11$, we find $\sin\alpha \lesssim 0.36$ at the 2$\sigma$ level (see e.g. also~\cite{Robens:2015gla}).
(ii) it induces Higgs like couplings of $\varphi$ to all SM particles proportional to $\sin\alpha$. Searches for $\varphi \to WW / ZZ$ lead to constraints on $\sin\alpha$ of the order of $\sim 0.3$ for $\varphi$ masses in the few hundred GeV range~\cite{Khachatryan:2015cwa,Aad:2015kna,Robens:2015gla}.
(iii) it leads to non-standard flavor violating couplings of $h$
\begin{eqnarray}
\Gamma_{\mu\tau (\tau\mu)}^h &\simeq& \sin\alpha\frac{2 v v_\phi }{M_E M_L} Y_{LE}\lambda_{\mu L(E)}\lambda_{\tau E(L)}\, .\label{Hcouplings}
\end{eqnarray}
Note that the flavor violating Higgs couplings scale as $v^2 v_\phi^2 /(m_S^2 M^2)$, with $M = M_E, M_L$ and $m_S=m_h$ ($m_S=m_\phi$) if $m_\phi\ll m_h$ ($m_h\ll m_\phi$). The two powers of $v_\phi$ are required to compensate the change in the $L_\mu - L_\tau$ charge by two units going from $\tau \to \mu$. An example diagram showing the leading order in $v$ and $v_\phi$ in the unbroken phase is shown in Fig.~\ref{htaumu_diagram}.

\bigskip
{\em Phenomenology.}
%
Using the flavor violating Higgs couplings in~\eq{Hcouplings}, the $h\to\tau\mu$ branching ratio is given (at leading order in $v$ and $v_\phi$) by
\begin{eqnarray}
&& \text{BR}(h \to \tau\mu)\simeq \frac{m_h}{2\pi \Gamma_h}\frac{v^2 v_\phi^2}{M_E^2M_L^2} \sin^2\alpha \nonumber \\
&& ~~~~~\times \left(Y_{LE}^2\lambda_{\mu L}^2 \lambda_{\tau E}^2  + Y_{LE}^2\lambda_{\tau L}^2\lambda_{\mu E}^2 \right)\,, ~~~~~
\end{eqnarray}
with the total Higgs width $\Gamma_h \simeq \Gamma_h^\text{SM} \cos^2\alpha $ and $\Gamma_h^\text{SM} \sim 4.1\,{\rm MeV}$~\cite{Heinemeyer:2013tqa}. We find that for couplings of $\mathcal O(1)$ and $v_\phi$ of the order of the electroweak scale, one can reach \% level branching ratios with vector-like lepton masses in the few TeV range.

Also the flavor conserving Higgs decay $h \to \mu\mu$ is modified significantly. We find
\begin{equation}
\frac{\text{BR}(h \to \mu\mu)}{\text{BR}(h \to \mu\mu)_\text{SM}} \simeq \left| 1 + \frac{v}{m_\mu} \tan\alpha \frac{2 v v_\phi}{M_E M_L} Y_{LE} \lambda_{\mu L} \lambda_{\mu E}  \right|^2 \,,
\end{equation}
For a \% level $h \to \tau\mu$ rate, we generically expect $O(1)$ corrections to the $h \to \mu\mu$ decay. The current bound of BR$(h\to\mu\mu) < 1.6 \times 10^{-3}$~\cite{Khachatryan:2014aep} already starts to constrain parts of the relevant parameter space.

\begin{figure*}[tbh]
\centering
\includegraphics[width=0.32\textwidth]{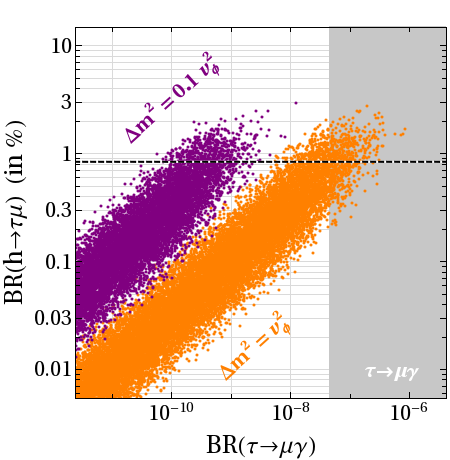} ~
\includegraphics[width=0.32\textwidth]{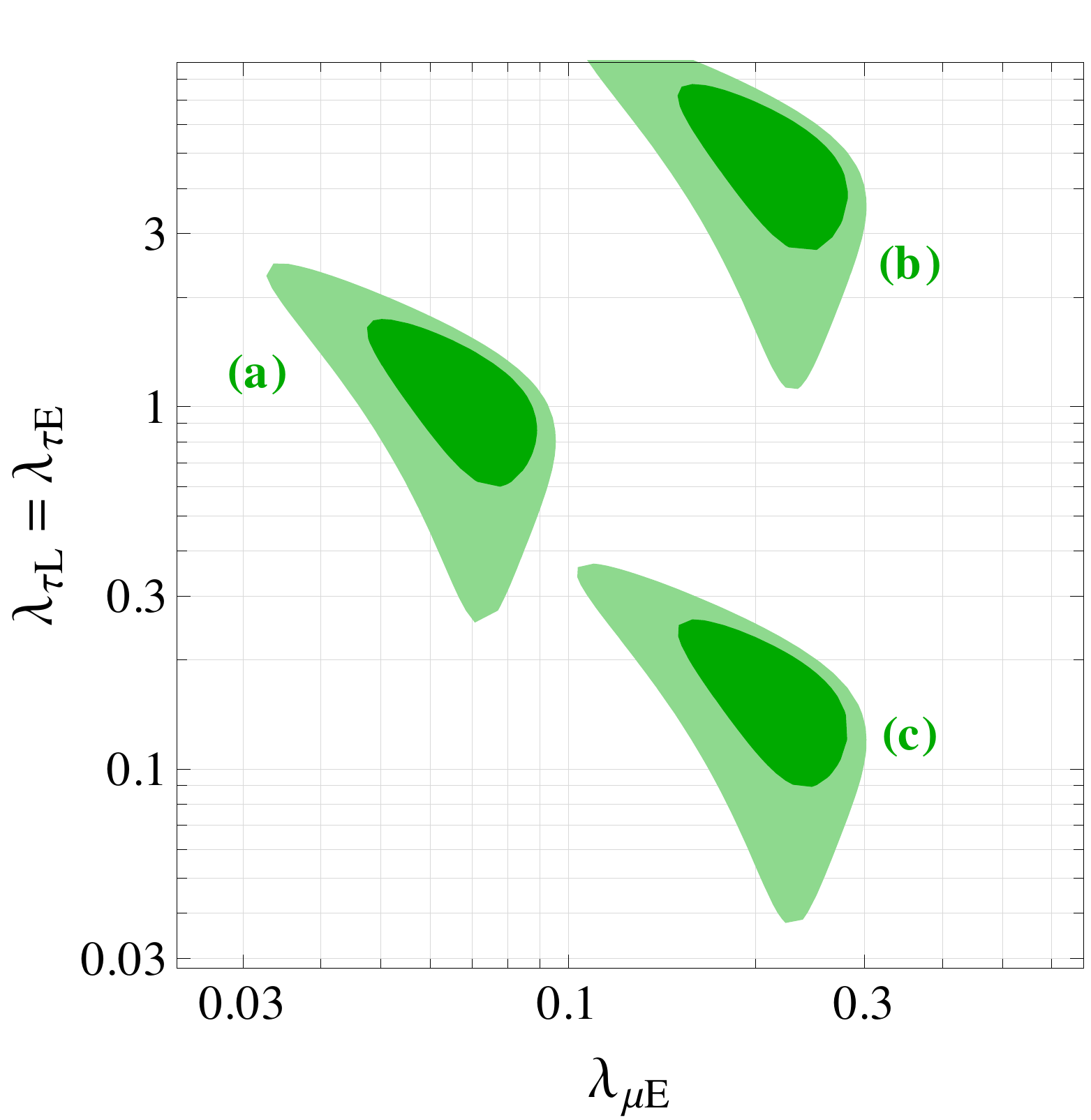} ~
\includegraphics[width=0.32\textwidth]{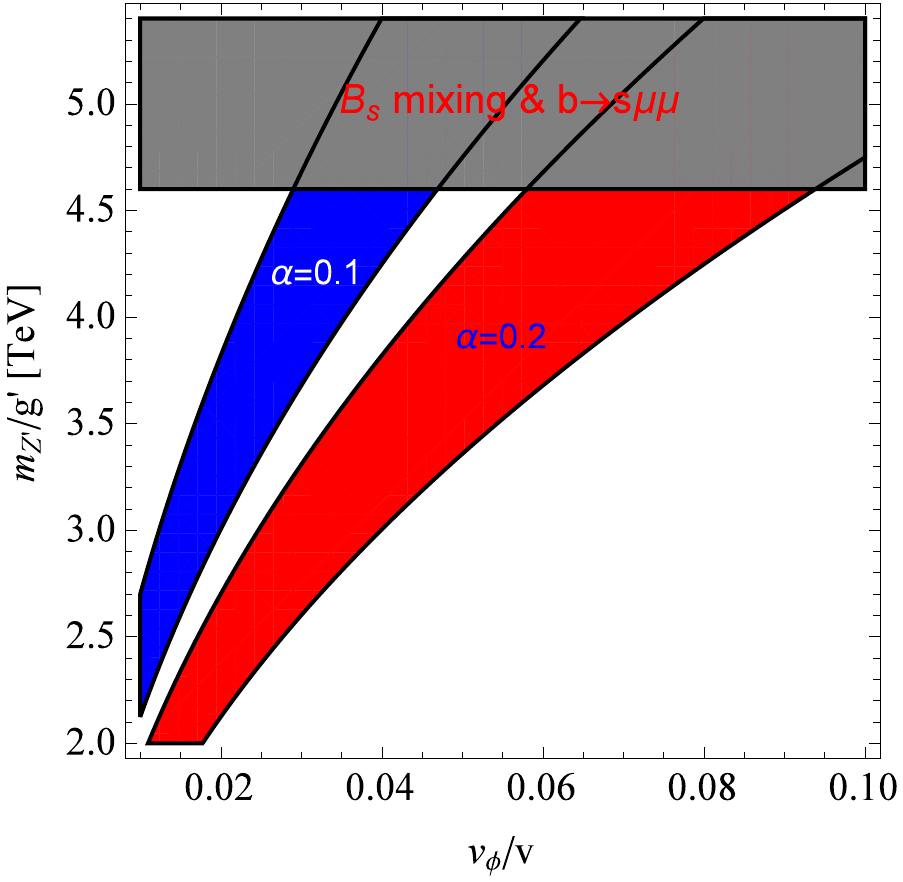}
\caption{Left: Correlations between $h\to\tau\mu$ and $\tau\to\mu\gamma$ for $\sin\alpha = 0.2$. The couplings $Y_{EL}, Y_{LE}, \lambda_{\mu L}, \lambda_{\mu E}, \lambda_{\tau L}, \lambda_{\tau E}$ are scanned in the range $0.5 - 2$ and and $M_E = M_L$ in the range $1~\text{TeV} - 3 ~\text{TeV}$ as well as $0.1 v < v_\phi < 2 v$. The gray region is excluded by the current bound on BR$(\tau\to \mu\gamma)$. The horizontal dashed line indicates the experimental central value of BR$(h\to\tau\mu)$.
Center: Regions in the $\lambda_{\mu E}$ vs. $\lambda_{\tau E} = \lambda_{\tau L} $ plane where both the $h \to \tau \mu$ signal and the $(g-2)_\mu$ discrepancy can be explained simultaneously. In the plot $M_E = M_L = 1$~TeV, $Y_{LE} = Y_{EL} = 2$, $\tan\alpha = 0.2$ and we consider 3 scenarios specified in the text.    
Right: Allowed regions in the $v_\phi/v$ vs. $m_{Z^\prime}/g^\prime$ plane from $h\to\tau\mu$ ($1\,\sigma$) and $\tau\to3\mu$ (95\% C.L.) for $\alpha=0.1$ (blue) and $\alpha=0.2$ (red). In the gray region the $b\to s\mu\mu$ data cannot be explained by NP without violating $B_s$ mixing bounds.}
\label{plot}
\end{figure*}

Compared to $h\to\tau\mu$, the flavor violating tau decay $\tau\to \mu\gamma$ shows a different decoupling with the NP scales. The dominant contributions to $\tau\to \mu\gamma$ come from loops of vector-like leptons together with the scalar $\varphi$ and pseudoscalar $a$. As $\tau \to \mu \gamma$ violates $L_\mu - L_\tau$ by two units, the $\varphi$ and $a$ loops cancel up to terms of order $\Delta m^2/M^2 = (m_\varphi^2 - m_a^2)/M^2$ where $M$ is the generic mass of the vector-like leptons. The middle diagram in Fig.~\ref{htaumu_diagram} shows an example diagram corresponding to the relevant contribution in the unbroken phase. We find the following $\tau\to\mu\gamma$ branching ratio
\begin{eqnarray}
&& \frac{\text{BR}(\tau \to \mu \gamma)}{\text{BR}(\tau \to \mu \nu\nu)} \simeq \frac{3 \alpha_\text{e}}{8\pi} \frac{v^6 \Delta m^4}{m_\tau^2 M_E^4 M_L^4} \left( \lambda_{\tau L}^2 \lambda_{\mu E}^2 + \lambda_{\tau E}^2 \lambda_{\mu L}^2\right) \nonumber \\
&& ~~~~~~ \times \left[ Y_{EL} + Y_{LE} \left( \frac{M_E}{M_L} + \frac{M_L}{M_E} \right) \right]^2  \,,
\end{eqnarray}
where we used the $\tau \to \mu \nu\nu$ branching ratio as convenient normalization. Due to the suppression by the small mass difference $\Delta m^2$, the constraint from BR$(\tau\to\mu\gamma) < 4.4\times 10^{-8}$~\cite{Aubert:2009ag,Hayasaka:2007vc}, can be easily satisfied in regions of parameter space that lead to a percent level BR$(h\to\tau\mu)$. This is illustrated in the left plot of Fig.~\ref{plot}. We fix $\sin\alpha = 0.2$ for $\Delta m^2=0.1 v^2_\phi$ and $\Delta m^2=v^2_\phi$ and scan the couplings $Y_{EL}, Y_{LE}, \lambda_{\mu L}, \lambda_{\mu E}, \lambda_{\tau L}, \lambda_{\tau L}$ in the range $0.5 - 2$. We also scan $1~\text{TeV} < M_E = M_L < 3 ~\text{TeV}$ and $0.1 v < v_\phi < 2 v$. 

For the corrections to the flavor conserving observable $(g-2)_\mu$, the suppression by $\Delta m^2/M^2$ is absent. The dominant contributions arise again from loops of vector-like leptons with $\varphi$ and $a$. However, in contrast to $\tau\to \mu\gamma$, the $\varphi$ and $a$ loops add up constructively. An example diagram at leading order in $v$ is shown in Fig.~\ref{htaumu_diagram}. We find
\begin{equation}
\Delta a_\mu \simeq \frac{1}{8\pi^2} \frac{m_\mu v}{M_E M_L} Y_{LE} \lambda_{\mu L} \lambda_{\mu E}  \,.
\end{equation}
For vector-like lepton masses $M$ in the few TeV range and $\mathcal O(1)$ couplings, this is in the right ball park to explain the discrepancy in Eq.~(\ref{eq:g-2}). The center plot of Fig.~\ref{plot} shows the regions of parameter space in the $\lambda_{\mu E}$ vs. $\lambda_{\tau E} = \lambda_{\tau L} $ plane where both the $h \to \tau \mu$ signal and the $(g-2)_\mu$ discrepancy can be explained simultaneously. In the plot we set $M_E = M_L = 1$~TeV, $Y_{LE} = Y_{EL} = 2$, $\tan\alpha = 0.2$ and consider the 3 scenarios: (a) $\lambda_{\mu L} = \lambda_{\mu E}$ and $v_\phi = v$; (b) $\lambda_{\mu L} = 0.1 \lambda_{\mu E}$ and $v_\phi = 0.1 v$; (c) $\lambda_{\mu L} = 0.1\lambda_{\mu E}$ and $v_\phi = 3 v$.   
 
An additional important constraint in the discussed setup comes from the flavor violating tau decay $\tau \to 3 \mu$ that is induced by tree level exchange of the $Z^\prime$. We find a strong correlation of the $\tau \to 3 \mu$ and $h \to \tau\mu$ branching ratios
\begin{equation}
 \frac{\text{BR}(\tau\to3\mu)}{\text{BR}(\tau\to\mu\nu\nu)} \simeq \frac{2 v^2 v_\phi^2 (g^\prime)^4}{m_{Z^\prime}^4} \frac{1}{\tan^2\alpha} \frac{\text{BR}(h \to \tau\mu)}{\text{BR}(h \to \tau \tau)} \,,
\end{equation}
where we neglected small contributions to $\tau \to 3 \mu$ from $Z$ boson exchange. The current upper limit of the branching ratio is BR$(\tau \to 3 \mu) < 1.2 \times 10^{-8}$ at $90\%$~C.L.~\cite{Amhis:2014hma} and sensitivities down to branching ratios of the order of $10^{-9}$ seem feasible at Belle II~\cite{Aushev:2010bq}. The current bound already sets strong constraints on the ratio of $Z^\prime$ mass and $g^\prime$ gauge coupling
\begin{equation} \label{eq:bound}
 \frac{m_{Z^\prime}}{g^\prime} \gtrsim 17\,\text{TeV}~\left(\frac{v_\phi}{v}\right)^{\frac{1}{2}} \left(\frac{0.2}{\tan\alpha}\right)^{\frac{1}{2}} \left(\frac{\text{BR}(h\to\tau\mu)}{0.84\%}\right)^{\frac{1}{4}} \,.
\end{equation}
If the $Z^\prime$ is to explain the hints for NP in $b\to s\mu\mu$ (once vector-like quarks are added) without violating bounds from $B_s$ meson oscillations, the $Z^\prime$ parameter space is contrained: $m_{Z^\prime}/g^\prime \lesssim 5$~TeV~\cite{Altmannshofer:2014cfa,Crivellin:2015mga}. Given the stringent lower bound in~(\ref{eq:bound}), a simultaneous explanation of a $h \to \tau\mu$ signal and the $B$ decay anomalies is only possible if $v_\phi$ is well below the electroweak scale. This is illustrated in the right plot of Fig.~\ref{plot}. We explicitly checked that such small values of $v_\phi$ are compatible with an explanation of $h \to \tau\mu$ and $(g-2)_\mu$.

\bigskip
{\em Summary and Outlook.}
%
In this Letter we proposed a framework that can give rise to a $h\to\tau\mu$ rate at current experimental sensitivities without violating the strong constraints from other flavor violating $\tau \to \mu$ processes, in particular from the $\tau \to \mu\gamma$ decay.
Mixing of the SM muons and taus with heavy vector-like leptons leads to LFV effects, that are controlled by a gauged $L_\mu - L_\tau$ symmetry. The relative size of $\tau \to \mu \gamma$ and $h\to\tau\mu$ is determined by the mass of the vector-like leptons and the mass of a scalar $\phi$ that breaks $L_\mu - L_\tau$, such that BR$(\tau \to \mu \gamma)$/BR$(h \to \tau\mu) \propto m_\varphi^4 / M^4$.
Interestingly, despite the protection of $\tau\to\mu\gamma$, our $L_\mu - L_\tau$ model allows to explain the anomalous magnetic moment of the muon because the corresponding flavor conserving dipole operator is not protected by the $L_\mu - L_\tau$ symmetry. Similarly, also the flavor conserving decay of the Higgs into muons is not protected by $L_\mu - L_\tau$ and can show sizable deviations from SM prediction, testable at the LHC. One can even account for the observed discrepancies in $b\to s\mu^+\mu^-$ data once vector-like quarks are added to the model, leading to sizable rates for $\tau\to3\mu$ observable at Belle II.

The concept of our explicit $L_\mu-L_\tau$ flavor model can be easily generalized. In the absence of new sources of electroweak symmetry breaking, a $h \to \tau\mu$ rate at the \% level is generically in conflict with the stringent bounds from $\tau \to \mu \gamma$~\cite{Altmannshofer:2015esa}. Barring fine tuned cancellations, the corresponding upper bound on BR$(h\to\tau\mu) \lesssim 10^{-6}$ is four orders of magnitude below the current experimental sensitivities for $h\to\tau\mu$. Even though the setup proposed in this Letter does not contain new sources of electroweak symmetry breaking, the constraint from $\tau\to\mu\gamma$ can be avoided because a flavor symmetry controls the LFV transitions and enforces cancellations among different contributions. The relevant dimension six operators leading to $h\to\tau\mu$ and $\tau \to \mu \gamma$ decays are forbidden by the flavor symmetry and require additional insertions of a flavon field $\phi$ 
to compensate the flavor charge. 
If the NP sector is sufficiently complex, it can contain multiple scales and effects in $\tau \to \mu \gamma$ can be decoupled from $h\to\tau\mu$. Of course many other choices of flavor symmetries are possible and it would be for example interesting to combine our framework with models that use flavor symmetries at the electroweak scale to explain the hierarchical structure of the SM quarks and leptons~\cite{Bauer:2015fxa,Bauer:2015kzy,Huitu:2016pwk}. It would also be interesting to examine the possibility that the scalar $\phi$ (and/or $a$) is responsible for the 750 GeV excess in $\gamma\gamma$ searches observed by ATLAS and CMS~\cite{diphoton1,diphoton2}. This seems in general possible as the model already contains vector-like leptons and quarks which can induce couplings of the scalars to photons and gluons.

\vspace{2mm}

\acknowledgments{We thank 
Stefania~Gori and Michael Spira for useful discussions. A.~Crivellin is supported by a Ambizione fellowship of the Swiss National Science Foundation. We acknowledge the hospitality of the Aspen Center for Physics where this project was initiated. Fermilab is operated by Fermi Research Alliance, LLC under Contract No. DE-AC02- 07CH11359 with the United States Department of Energy.}

\bibliography{BIB}

\begin{thebibliography}{93}
\expandafter\ifx\csname natexlab\endcsname\relax\def\natexlab#1{#1}\fi
\expandafter\ifx\csname bibnamefont\endcsname\relax
  \def\bibnamefont#1{#1}\fi
\expandafter\ifx\csname bibfnamefont\endcsname\relax
  \def\bibfnamefont#1{#1}\fi
\expandafter\ifx\csname citenamefont\endcsname\relax
  \def\citenamefont#1{#1}\fi
\expandafter\ifx\csname url\endcsname\relax
  \def\url#1{\texttt{#1}}\fi
\expandafter\ifx\csname urlprefix\endcsname\relax\def\urlprefix{URL }\fi
\providecommand{\bibinfo}[2]{#2}
\providecommand{\eprint}[2][]{\url{#2}}

\bibitem[{\citenamefont{Davidson and Grenier}(2010)}]{Davidson:2010xv}
\bibinfo{author}{\bibfnamefont{S.}~\bibnamefont{Davidson}} \bibnamefont{and}
  \bibinfo{author}{\bibfnamefont{G.~J.} \bibnamefont{Grenier}},
  \bibinfo{journal}{Phys. Rev.} \textbf{\bibinfo{volume}{D81}},
  \bibinfo{pages}{095016} (\bibinfo{year}{2010}), \eprint{1001.0434}.

\bibitem[{\citenamefont{Blankenburg et~al.}(2012)\citenamefont{Blankenburg,
  Ellis, and Isidori}}]{Blankenburg:2012ex}
\bibinfo{author}{\bibfnamefont{G.}~\bibnamefont{Blankenburg}},
  \bibinfo{author}{\bibfnamefont{J.}~\bibnamefont{Ellis}}, \bibnamefont{and}
  \bibinfo{author}{\bibfnamefont{G.}~\bibnamefont{Isidori}},
  \bibinfo{journal}{Phys. Lett.} \textbf{\bibinfo{volume}{B712}},
  \bibinfo{pages}{386} (\bibinfo{year}{2012}), \eprint{1202.5704}.

\bibitem[{\citenamefont{Harnik et~al.}(2013)\citenamefont{Harnik, Kopp, and
  Zupan}}]{Harnik:2012pb}
\bibinfo{author}{\bibfnamefont{R.}~\bibnamefont{Harnik}},
  \bibinfo{author}{\bibfnamefont{J.}~\bibnamefont{Kopp}}, \bibnamefont{and}
  \bibinfo{author}{\bibfnamefont{J.}~\bibnamefont{Zupan}},
  \bibinfo{journal}{JHEP} \textbf{\bibinfo{volume}{03}}, \bibinfo{pages}{026}
  (\bibinfo{year}{2013}), \eprint{1209.1397}.

\bibitem[{\citenamefont{Khachatryan
  et~al.}(2015{\natexlab{a}})}]{Khachatryan:2015kon}
\bibinfo{author}{\bibfnamefont{V.}~\bibnamefont{Khachatryan}}
  \bibnamefont{et~al.} (\bibinfo{collaboration}{CMS}), \bibinfo{journal}{Phys.
  Lett.} \textbf{\bibinfo{volume}{B749}}, \bibinfo{pages}{337}
  (\bibinfo{year}{2015}{\natexlab{a}}),
  \eprint{\href{http://arxiv.org/abs/1502.07400}{arXiv:1502.07400 [hep-ex]}}.

\bibitem[{\citenamefont{Aad et~al.}(2016{\natexlab{a}})}]{Aad:2016blu}
\bibinfo{author}{\bibfnamefont{G.}~\bibnamefont{Aad}} \bibnamefont{et~al.}
  (\bibinfo{collaboration}{ATLAS}) (\bibinfo{year}{2016}{\natexlab{a}}),
  \eprint{1604.07730}.

\bibitem[{\citenamefont{Campos et~al.}(2015)\citenamefont{Campos,
  Hern{\'a}ndez, P{\"a}s, and Schumacher}}]{Campos:2014zaa}
\bibinfo{author}{\bibfnamefont{M.~D.} \bibnamefont{Campos}},
  \bibinfo{author}{\bibfnamefont{A.~E.~C.} \bibnamefont{Hern{\'a}ndez}},
  \bibinfo{author}{\bibfnamefont{H.}~\bibnamefont{P{\"a}s}}, \bibnamefont{and}
  \bibinfo{author}{\bibfnamefont{E.}~\bibnamefont{Schumacher}},
  \bibinfo{journal}{Phys. Rev.} \textbf{\bibinfo{volume}{D91}},
  \bibinfo{pages}{116011} (\bibinfo{year}{2015}),
  \eprint{\href{http://arxiv.org/abs/1408.1652}{[arXiv:1408.1652 [hep-ph]]}}.

\bibitem[{\citenamefont{Aristizabal~Sierra and Vicente}(2014)}]{Sierra:2014nqa}
\bibinfo{author}{\bibfnamefont{D.}~\bibnamefont{Aristizabal~Sierra}}
  \bibnamefont{and} \bibinfo{author}{\bibfnamefont{A.}~\bibnamefont{Vicente}},
  \bibinfo{journal}{Phys. Rev.} \textbf{\bibinfo{volume}{D90}},
  \bibinfo{pages}{115004} (\bibinfo{year}{2014}),
  \eprint{\href{http://arxiv.org/abs/1409.7690}{[arXiv:1409.7690 [hep-ph]]}}.

\bibitem[{\citenamefont{Heeck et~al.}(2015)\citenamefont{Heeck, Holthausen,
  Rodejohann, and Shimizu}}]{Heeck:2014qea}
\bibinfo{author}{\bibfnamefont{J.}~\bibnamefont{Heeck}},
  \bibinfo{author}{\bibfnamefont{M.}~\bibnamefont{Holthausen}},
  \bibinfo{author}{\bibfnamefont{W.}~\bibnamefont{Rodejohann}},
  \bibnamefont{and} \bibinfo{author}{\bibfnamefont{Y.}~\bibnamefont{Shimizu}},
  \bibinfo{journal}{Nucl. Phys.} \textbf{\bibinfo{volume}{B896}},
  \bibinfo{pages}{281} (\bibinfo{year}{2015}),
  \eprint{\href{http://arxiv.org/abs/1412.3671}{[arXiv:1412.3671 [hep-ph]]}}.

\bibitem[{\citenamefont{Crivellin
  et~al.}(2015{\natexlab{a}})\citenamefont{Crivellin, D'Ambrosio, and
  Heeck}}]{Crivellin:2015mga}
\bibinfo{author}{\bibfnamefont{A.}~\bibnamefont{Crivellin}},
  \bibinfo{author}{\bibfnamefont{G.}~\bibnamefont{D'Ambrosio}},
  \bibnamefont{and} \bibinfo{author}{\bibfnamefont{J.}~\bibnamefont{Heeck}},
  \bibinfo{journal}{Phys. Rev. Lett.} \textbf{\bibinfo{volume}{114}},
  \bibinfo{pages}{151801} (\bibinfo{year}{2015}{\natexlab{a}}),
  \eprint{\href{http://arxiv.org/abs/1501.00993}{[arXiv:1501.00993 [hep-ph]]}}.

\bibitem[{\citenamefont{Dorsner et~al.}(2015)\citenamefont{Dorsner, Fajfer,
  Greljo, Kamenik, Kosnik et~al.}}]{Dorsner:2015mja}
\bibinfo{author}{\bibfnamefont{I.}~\bibnamefont{Dorsner}},
  \bibinfo{author}{\bibfnamefont{S.}~\bibnamefont{Fajfer}},
  \bibinfo{author}{\bibfnamefont{A.}~\bibnamefont{Greljo}},
  \bibinfo{author}{\bibfnamefont{J.~F.} \bibnamefont{Kamenik}},
  \bibinfo{author}{\bibfnamefont{N.}~\bibnamefont{Kosnik}},
  \bibnamefont{et~al.}, \bibinfo{journal}{JHEP}
  \textbf{\bibinfo{volume}{1506}}, \bibinfo{pages}{108} (\bibinfo{year}{2015}),
  \eprint{\href{http://arxiv.org/abs/1502.07784}{[arXiv:1502.07784 [hep-ph]]}}.

\bibitem[{\citenamefont{Omura et~al.}(2015)\citenamefont{Omura, Senaha, and
  Tobe}}]{Omura:2015nja}
\bibinfo{author}{\bibfnamefont{Y.}~\bibnamefont{Omura}},
  \bibinfo{author}{\bibfnamefont{E.}~\bibnamefont{Senaha}}, \bibnamefont{and}
  \bibinfo{author}{\bibfnamefont{K.}~\bibnamefont{Tobe}},
  \bibinfo{journal}{JHEP} \textbf{\bibinfo{volume}{1505}}, \bibinfo{pages}{028}
  (\bibinfo{year}{2015}),
  \eprint{\href{http://arxiv.org/abs/1502.07824}{[arXiv:1502.07824 [hep-ph]]}}.

\bibitem[{\citenamefont{Crivellin
  et~al.}(2015{\natexlab{b}})\citenamefont{Crivellin, D'Ambrosio, and
  Heeck}}]{Crivellin:2015lwa}
\bibinfo{author}{\bibfnamefont{A.}~\bibnamefont{Crivellin}},
  \bibinfo{author}{\bibfnamefont{G.}~\bibnamefont{D'Ambrosio}},
  \bibnamefont{and} \bibinfo{author}{\bibfnamefont{J.}~\bibnamefont{Heeck}},
  \bibinfo{journal}{Phys. Rev.} \textbf{\bibinfo{volume}{D91}},
  \bibinfo{pages}{075006} (\bibinfo{year}{2015}{\natexlab{b}}),
  \eprint{\href{http://arxiv.org/abs/1503.03477}{[arXiv:1503.03477 [hep-ph]]}}.

\bibitem[{\citenamefont{Chiang et~al.}(2015)\citenamefont{Chiang, Fukuda,
  Takeuchi, and Yanagida}}]{Chiang:2015cba}
\bibinfo{author}{\bibfnamefont{C.-W.} \bibnamefont{Chiang}},
  \bibinfo{author}{\bibfnamefont{H.}~\bibnamefont{Fukuda}},
  \bibinfo{author}{\bibfnamefont{M.}~\bibnamefont{Takeuchi}}, \bibnamefont{and}
  \bibinfo{author}{\bibfnamefont{T.~T.} \bibnamefont{Yanagida}}
  (\bibinfo{year}{2015}), \eprint{1507.04354}.

\bibitem[{\citenamefont{Altmannshofer et~al.}(2015)\citenamefont{Altmannshofer,
  Gori, Kagan, Silvestrini, and Zupan}}]{Altmannshofer:2015esa}
\bibinfo{author}{\bibfnamefont{W.}~\bibnamefont{Altmannshofer}},
  \bibinfo{author}{\bibfnamefont{S.}~\bibnamefont{Gori}},
  \bibinfo{author}{\bibfnamefont{A.~L.} \bibnamefont{Kagan}},
  \bibinfo{author}{\bibfnamefont{L.}~\bibnamefont{Silvestrini}},
  \bibnamefont{and} \bibinfo{author}{\bibfnamefont{J.}~\bibnamefont{Zupan}}
  (\bibinfo{year}{2015}), \eprint{1507.07927}.

\bibitem[{\citenamefont{Aloni et~al.}(2015)\citenamefont{Aloni, Nir, and
  Stamou}}]{Aloni:2015wvn}
\bibinfo{author}{\bibfnamefont{D.}~\bibnamefont{Aloni}},
  \bibinfo{author}{\bibfnamefont{Y.}~\bibnamefont{Nir}}, \bibnamefont{and}
  \bibinfo{author}{\bibfnamefont{E.}~\bibnamefont{Stamou}}
  (\bibinfo{year}{2015}), \eprint{1511.00979}.

\bibitem[{\citenamefont{Alvarado et~al.}(2016)\citenamefont{Alvarado,
  Capdevilla, Delgado, and Martin}}]{Alvarado:2016par}
\bibinfo{author}{\bibfnamefont{C.}~\bibnamefont{Alvarado}},
  \bibinfo{author}{\bibfnamefont{R.~M.} \bibnamefont{Capdevilla}},
  \bibinfo{author}{\bibfnamefont{A.}~\bibnamefont{Delgado}}, \bibnamefont{and}
  \bibinfo{author}{\bibfnamefont{A.}~\bibnamefont{Martin}}
  (\bibinfo{year}{2016}), \eprint{1602.08506}.

\bibitem[{\citenamefont{Amhis et~al.}(2014)}]{Amhis:2014hma}
\bibinfo{author}{\bibfnamefont{Y.}~\bibnamefont{Amhis}} \bibnamefont{et~al.}
  (\bibinfo{collaboration}{Heavy Flavor Averaging Group (HFAG)})
  (\bibinfo{year}{2014}),
  \eprint{\href{http://arxiv.org/abs/1412.7515}{arXiv:1412.7515 [hep-ex]}}.

\bibitem[{\citenamefont{Aaij et~al.}(2014)}]{Aaij:2014ora}
\bibinfo{author}{\bibfnamefont{R.}~\bibnamefont{Aaij}} \bibnamefont{et~al.}
  (\bibinfo{collaboration}{LHCb}), \bibinfo{journal}{Phys. Rev. Lett.}
  \textbf{\bibinfo{volume}{113}}, \bibinfo{pages}{151601}
  (\bibinfo{year}{2014}), \eprint{1406.6482}.

\bibitem[{\citenamefont{Bobeth et~al.}(2007)\citenamefont{Bobeth, Hiller, and
  Piranishvili}}]{Bobeth:2007dw}
\bibinfo{author}{\bibfnamefont{C.}~\bibnamefont{Bobeth}},
  \bibinfo{author}{\bibfnamefont{G.}~\bibnamefont{Hiller}}, \bibnamefont{and}
  \bibinfo{author}{\bibfnamefont{G.}~\bibnamefont{Piranishvili}},
  \bibinfo{journal}{JHEP} \textbf{\bibinfo{volume}{12}}, \bibinfo{pages}{040}
  (\bibinfo{year}{2007}), \eprint{0709.4174}.

\bibitem[{\citenamefont{Altmannshofer and
  Straub}(2015{\natexlab{a}})}]{Altmannshofer:2014rta}
\bibinfo{author}{\bibfnamefont{W.}~\bibnamefont{Altmannshofer}}
  \bibnamefont{and} \bibinfo{author}{\bibfnamefont{D.~M.}
  \bibnamefont{Straub}}, \bibinfo{journal}{Eur. Phys. J.}
  \textbf{\bibinfo{volume}{C75}}, \bibinfo{pages}{382}
  (\bibinfo{year}{2015}{\natexlab{a}}), \eprint{1411.3161}.

\bibitem[{\citenamefont{Altmannshofer and
  Straub}(2015{\natexlab{b}})}]{Altmannshofer:2015sma}
\bibinfo{author}{\bibfnamefont{W.}~\bibnamefont{Altmannshofer}}
  \bibnamefont{and} \bibinfo{author}{\bibfnamefont{D.~M.} \bibnamefont{Straub}}
  (\bibinfo{year}{2015}{\natexlab{b}}), \eprint{1503.06199}.

\bibitem[{\citenamefont{Descotes-Genon
  et~al.}(2015)\citenamefont{Descotes-Genon, Hofer, Matias, and
  Virto}}]{Descotes-Genon:2015uva}
\bibinfo{author}{\bibfnamefont{S.}~\bibnamefont{Descotes-Genon}},
  \bibinfo{author}{\bibfnamefont{L.}~\bibnamefont{Hofer}},
  \bibinfo{author}{\bibfnamefont{J.}~\bibnamefont{Matias}}, \bibnamefont{and}
  \bibinfo{author}{\bibfnamefont{J.}~\bibnamefont{Virto}}
  (\bibinfo{year}{2015}), \eprint{1510.04239}.

\bibitem[{\citenamefont{Hurth et~al.}(2016)\citenamefont{Hurth, Mahmoudi, and
  Neshatpour}}]{Hurth:2016fbr}
\bibinfo{author}{\bibfnamefont{T.}~\bibnamefont{Hurth}},
  \bibinfo{author}{\bibfnamefont{F.}~\bibnamefont{Mahmoudi}}, \bibnamefont{and}
  \bibinfo{author}{\bibfnamefont{S.}~\bibnamefont{Neshatpour}}
  (\bibinfo{year}{2016}), \eprint{1603.00865}.

\bibitem[{\citenamefont{Lyon and Zwicky}(2014)}]{Lyon:2014hpa}
\bibinfo{author}{\bibfnamefont{J.}~\bibnamefont{Lyon}} \bibnamefont{and}
  \bibinfo{author}{\bibfnamefont{R.}~\bibnamefont{Zwicky}}
  (\bibinfo{year}{2014}), \eprint{1406.0566}.

\bibitem[{\citenamefont{Descotes-Genon
  et~al.}(2014)\citenamefont{Descotes-Genon, Hofer, Matias, and
  Virto}}]{Descotes-Genon:2014uoa}
\bibinfo{author}{\bibfnamefont{S.}~\bibnamefont{Descotes-Genon}},
  \bibinfo{author}{\bibfnamefont{L.}~\bibnamefont{Hofer}},
  \bibinfo{author}{\bibfnamefont{J.}~\bibnamefont{Matias}}, \bibnamefont{and}
  \bibinfo{author}{\bibfnamefont{J.}~\bibnamefont{Virto}},
  \bibinfo{journal}{JHEP} \textbf{\bibinfo{volume}{12}}, \bibinfo{pages}{125}
  (\bibinfo{year}{2014}), \eprint{1407.8526}.

\bibitem[{\citenamefont{Jäger and Martin~Camalich}(2016)}]{Jager:2014rwa}
\bibinfo{author}{\bibfnamefont{S.}~\bibnamefont{Jäger}} \bibnamefont{and}
  \bibinfo{author}{\bibfnamefont{J.}~\bibnamefont{Martin~Camalich}},
  \bibinfo{journal}{Phys. Rev.} \textbf{\bibinfo{volume}{D93}},
  \bibinfo{pages}{014028} (\bibinfo{year}{2016}), \eprint{1412.3183}.

\bibitem[{\citenamefont{Altmannshofer
  et~al.}(2014{\natexlab{a}})\citenamefont{Altmannshofer, Gori, Pospelov, and
  Yavin}}]{Altmannshofer:2014cfa}
\bibinfo{author}{\bibfnamefont{W.}~\bibnamefont{Altmannshofer}},
  \bibinfo{author}{\bibfnamefont{S.}~\bibnamefont{Gori}},
  \bibinfo{author}{\bibfnamefont{M.}~\bibnamefont{Pospelov}}, \bibnamefont{and}
  \bibinfo{author}{\bibfnamefont{I.}~\bibnamefont{Yavin}},
  \bibinfo{journal}{Phys. Rev.} \textbf{\bibinfo{volume}{D89}},
  \bibinfo{pages}{095033} (\bibinfo{year}{2014}{\natexlab{a}}),
  \eprint{\href{http://arxiv.org/abs/1403.1269}{[arXiv:1403.1269 [hep-ph]]}}.

\bibitem[{\citenamefont{Altmannshofer and Yavin}(2015)}]{Altmannshofer:2015mqa}
\bibinfo{author}{\bibfnamefont{W.}~\bibnamefont{Altmannshofer}}
  \bibnamefont{and} \bibinfo{author}{\bibfnamefont{I.}~\bibnamefont{Yavin}},
  \bibinfo{journal}{Phys. Rev.} \textbf{\bibinfo{volume}{D92}},
  \bibinfo{pages}{075022} (\bibinfo{year}{2015}), \eprint{1508.07009}.

\bibitem[{\citenamefont{Niehoff et~al.}(2015)\citenamefont{Niehoff, Stangl, and
  Straub}}]{Niehoff:2015bfa}
\bibinfo{author}{\bibfnamefont{C.}~\bibnamefont{Niehoff}},
  \bibinfo{author}{\bibfnamefont{P.}~\bibnamefont{Stangl}}, \bibnamefont{and}
  \bibinfo{author}{\bibfnamefont{D.~M.} \bibnamefont{Straub}},
  \bibinfo{journal}{Phys. Lett.} \textbf{\bibinfo{volume}{B747}},
  \bibinfo{pages}{182} (\bibinfo{year}{2015}), \eprint{1503.03865}.

\bibitem[{\citenamefont{Aristizabal~Sierra
  et~al.}(2015)\citenamefont{Aristizabal~Sierra, Staub, and
  Vicente}}]{Sierra:2015fma}
\bibinfo{author}{\bibfnamefont{D.}~\bibnamefont{Aristizabal~Sierra}},
  \bibinfo{author}{\bibfnamefont{F.}~\bibnamefont{Staub}}, \bibnamefont{and}
  \bibinfo{author}{\bibfnamefont{A.}~\bibnamefont{Vicente}},
  \bibinfo{journal}{Phys. Rev.} \textbf{\bibinfo{volume}{D92}},
  \bibinfo{pages}{015001} (\bibinfo{year}{2015}), \eprint{1503.06077}.

\bibitem[{\citenamefont{Celis et~al.}(2015)\citenamefont{Celis, Fuentes-Martin,
  Jung, and Serodio}}]{Celis:2015ara}
\bibinfo{author}{\bibfnamefont{A.}~\bibnamefont{Celis}},
  \bibinfo{author}{\bibfnamefont{J.}~\bibnamefont{Fuentes-Martin}},
  \bibinfo{author}{\bibfnamefont{M.}~\bibnamefont{Jung}}, \bibnamefont{and}
  \bibinfo{author}{\bibfnamefont{H.}~\bibnamefont{Serodio}},
  \bibinfo{journal}{Phys. Rev.} \textbf{\bibinfo{volume}{D92}},
  \bibinfo{pages}{015007} (\bibinfo{year}{2015}), \eprint{1505.03079}.

\bibitem[{\citenamefont{Bélanger et~al.}(2015)\citenamefont{Bélanger,
  Delaunay, and Westhoff}}]{Belanger:2015nma}
\bibinfo{author}{\bibfnamefont{G.}~\bibnamefont{Bélanger}},
  \bibinfo{author}{\bibfnamefont{C.}~\bibnamefont{Delaunay}}, \bibnamefont{and}
  \bibinfo{author}{\bibfnamefont{S.}~\bibnamefont{Westhoff}},
  \bibinfo{journal}{Phys. Rev.} \textbf{\bibinfo{volume}{D92}},
  \bibinfo{pages}{055021} (\bibinfo{year}{2015}), \eprint{1507.06660}.

\bibitem[{\citenamefont{Falkowski et~al.}(2015)\citenamefont{Falkowski,
  Nardecchia, and Ziegler}}]{Falkowski:2015zwa}
\bibinfo{author}{\bibfnamefont{A.}~\bibnamefont{Falkowski}},
  \bibinfo{author}{\bibfnamefont{M.}~\bibnamefont{Nardecchia}},
  \bibnamefont{and} \bibinfo{author}{\bibfnamefont{R.}~\bibnamefont{Ziegler}},
  \bibinfo{journal}{JHEP} \textbf{\bibinfo{volume}{11}}, \bibinfo{pages}{173}
  (\bibinfo{year}{2015}), \eprint{1509.01249}.

\bibitem[{\citenamefont{Carmona and Goertz}(2015)}]{Carmona:2015ena}
\bibinfo{author}{\bibfnamefont{A.}~\bibnamefont{Carmona}} \bibnamefont{and}
  \bibinfo{author}{\bibfnamefont{F.}~\bibnamefont{Goertz}}
  (\bibinfo{year}{2015}), \eprint{1510.07658}.

\bibitem[{\citenamefont{Buras and De~Fazio}(2016)}]{Buras:2016dxz}
\bibinfo{author}{\bibfnamefont{A.~J.} \bibnamefont{Buras}} \bibnamefont{and}
  \bibinfo{author}{\bibfnamefont{F.}~\bibnamefont{De~Fazio}}
  (\bibinfo{year}{2016}), \eprint{1604.02344}.

\bibitem[{\citenamefont{Gripaios et~al.}(2015)\citenamefont{Gripaios,
  Nardecchia, and Renner}}]{Gripaios:2014tna}
\bibinfo{author}{\bibfnamefont{B.}~\bibnamefont{Gripaios}},
  \bibinfo{author}{\bibfnamefont{M.}~\bibnamefont{Nardecchia}},
  \bibnamefont{and} \bibinfo{author}{\bibfnamefont{S.~A.}
  \bibnamefont{Renner}}, \bibinfo{journal}{JHEP} \textbf{\bibinfo{volume}{05}},
  \bibinfo{pages}{006} (\bibinfo{year}{2015}), \eprint{1412.1791}.

\bibitem[{\citenamefont{Bečirević et~al.}(2015)\citenamefont{Bečirević,
  Fajfer, and Košnik}}]{Becirevic:2015asa}
\bibinfo{author}{\bibfnamefont{D.}~\bibnamefont{Bečirević}},
  \bibinfo{author}{\bibfnamefont{S.}~\bibnamefont{Fajfer}}, \bibnamefont{and}
  \bibinfo{author}{\bibfnamefont{N.}~\bibnamefont{Košnik}},
  \bibinfo{journal}{Phys. Rev.} \textbf{\bibinfo{volume}{D92}},
  \bibinfo{pages}{014016} (\bibinfo{year}{2015}), \eprint{1503.09024}.

\bibitem[{\citenamefont{de~Medeiros~Varzielas and
  Hiller}(2015)}]{Varzielas:2015iva}
\bibinfo{author}{\bibfnamefont{I.}~\bibnamefont{de~Medeiros~Varzielas}}
  \bibnamefont{and} \bibinfo{author}{\bibfnamefont{G.}~\bibnamefont{Hiller}},
  \bibinfo{journal}{JHEP} \textbf{\bibinfo{volume}{06}}, \bibinfo{pages}{072}
  (\bibinfo{year}{2015}), \eprint{1503.01084}.

\bibitem[{\citenamefont{Alonso et~al.}(2015)\citenamefont{Alonso, Grinstein,
  and Camalich}}]{Alonso:2015sja}
\bibinfo{author}{\bibfnamefont{R.}~\bibnamefont{Alonso}},
  \bibinfo{author}{\bibfnamefont{B.}~\bibnamefont{Grinstein}},
  \bibnamefont{and} \bibinfo{author}{\bibfnamefont{J.~M.}
  \bibnamefont{Camalich}} (\bibinfo{year}{2015}),
  \eprint{\href{http://arxiv.org/abs/1505.05164}{arXiv:1505.05164 [hep-ph]}}.

\bibitem[{\citenamefont{Calibbi et~al.}(2015)\citenamefont{Calibbi, Crivellin,
  and Ota}}]{Calibbi:2015kma}
\bibinfo{author}{\bibfnamefont{L.}~\bibnamefont{Calibbi}},
  \bibinfo{author}{\bibfnamefont{A.}~\bibnamefont{Crivellin}},
  \bibnamefont{and} \bibinfo{author}{\bibfnamefont{T.}~\bibnamefont{Ota}},
  \bibinfo{journal}{Phys. Rev. Lett.} \textbf{\bibinfo{volume}{115}},
  \bibinfo{pages}{181801} (\bibinfo{year}{2015}),
  \eprint{\href{http://arxiv.org/abs/1506.02661}{arXiv:1506.02661 [hep-ph]}}.

\bibitem[{\citenamefont{Bauer and Neubert}(2016)}]{Bauer:2015knc}
\bibinfo{author}{\bibfnamefont{M.}~\bibnamefont{Bauer}} \bibnamefont{and}
  \bibinfo{author}{\bibfnamefont{M.}~\bibnamefont{Neubert}},
  \bibinfo{journal}{Phys. Rev. Lett.} \textbf{\bibinfo{volume}{116}},
  \bibinfo{pages}{141802} (\bibinfo{year}{2016}), \eprint{1511.01900}.

\bibitem[{\citenamefont{Doršner et~al.}(2016)\citenamefont{Doršner, Fajfer,
  Greljo, Kamenik, and Košnik}}]{Dorsner:2016wpm}
\bibinfo{author}{\bibfnamefont{I.}~\bibnamefont{Doršner}},
  \bibinfo{author}{\bibfnamefont{S.}~\bibnamefont{Fajfer}},
  \bibinfo{author}{\bibfnamefont{A.}~\bibnamefont{Greljo}},
  \bibinfo{author}{\bibfnamefont{J.~F.} \bibnamefont{Kamenik}},
  \bibnamefont{and} \bibinfo{author}{\bibfnamefont{N.}~\bibnamefont{Košnik}}
  (\bibinfo{year}{2016}), \eprint{1603.04993}.

\bibitem[{\citenamefont{Bennett et~al.}(2006)}]{Bennett2006}
\bibinfo{author}{\bibfnamefont{G.}~\bibnamefont{Bennett}} \bibnamefont{et~al.}
  (\bibinfo{collaboration}{Muon $(g-2)$ Collaboration}),
  \bibinfo{journal}{Phys. Rev.} \textbf{\bibinfo{volume}{D73}},
  \bibinfo{pages}{072003} (\bibinfo{year}{2006}),
  \eprint{\href{http://arXiv.org/abs/hep-ex/0602035}{[arXiv:hep-ex/0602035]}}.

\bibitem[{\citenamefont{Olive et~al.}(2014)}]{Olive2014}
\bibinfo{author}{\bibfnamefont{K.}~\bibnamefont{Olive}} \bibnamefont{et~al.}
  (\bibinfo{collaboration}{Particle Data Group}), \bibinfo{journal}{Chin.
  Phys.} \textbf{\bibinfo{volume}{C38}}, \bibinfo{pages}{090001}
  (\bibinfo{year}{2014}).

\bibitem[{\citenamefont{Aoyama et~al.}(2012)\citenamefont{Aoyama, Hayakawa,
  Kinoshita, and Nio}}]{Aoyama2012}
\bibinfo{author}{\bibfnamefont{T.}~\bibnamefont{Aoyama}},
  \bibinfo{author}{\bibfnamefont{M.}~\bibnamefont{Hayakawa}},
  \bibinfo{author}{\bibfnamefont{T.}~\bibnamefont{Kinoshita}},
  \bibnamefont{and} \bibinfo{author}{\bibfnamefont{M.}~\bibnamefont{Nio}},
  \bibinfo{journal}{Phys. Rev. Lett.} \textbf{\bibinfo{volume}{109}},
  \bibinfo{pages}{111808} (\bibinfo{year}{2012}),
  \eprint{\href{http://arxiv.org/abs/1205.5370}{[arXiv:1205.5370 [hep-ph]]}}.

\bibitem[{\citenamefont{Czarnecki et~al.}(1995)\citenamefont{Czarnecki, Krause,
  and Marciano}}]{Czarnecki1995}
\bibinfo{author}{\bibfnamefont{A.}~\bibnamefont{Czarnecki}},
  \bibinfo{author}{\bibfnamefont{B.}~\bibnamefont{Krause}}, \bibnamefont{and}
  \bibinfo{author}{\bibfnamefont{W.~J.} \bibnamefont{Marciano}},
  \bibinfo{journal}{Phys. Rev.} \textbf{\bibinfo{volume}{D52}},
  \bibinfo{pages}{2619} (\bibinfo{year}{1995}),
  \eprint{\href{http://arxiv.org/abs/hep-ph/9506256}{[arXiv:hep-ph/9506256]}}.

\bibitem[{\citenamefont{Czarnecki et~al.}(1996)\citenamefont{Czarnecki, Krause,
  and Marciano}}]{Czarnecki1996}
\bibinfo{author}{\bibfnamefont{A.}~\bibnamefont{Czarnecki}},
  \bibinfo{author}{\bibfnamefont{B.}~\bibnamefont{Krause}}, \bibnamefont{and}
  \bibinfo{author}{\bibfnamefont{W.~J.} \bibnamefont{Marciano}},
  \bibinfo{journal}{Phys. Rev. Lett.} \textbf{\bibinfo{volume}{76}},
  \bibinfo{pages}{3267} (\bibinfo{year}{1996}),
  \eprint{\href{http://arxiv.org/abs/hep-ph/9512369}{[arXiv:hep-ph/9512369]}}.

\bibitem[{\citenamefont{Gnendiger et~al.}(2013)\citenamefont{Gnendiger,
  St\"ockinger, and St\"ockinger-Kim}}]{Gnendiger2013}
\bibinfo{author}{\bibfnamefont{C.}~\bibnamefont{Gnendiger}},
  \bibinfo{author}{\bibfnamefont{D.}~\bibnamefont{St\"ockinger}},
  \bibnamefont{and}
  \bibinfo{author}{\bibfnamefont{H.}~\bibnamefont{St\"ockinger-Kim}},
  \bibinfo{journal}{Phys. Rev.} \textbf{\bibinfo{volume}{D88}},
  \bibinfo{pages}{053005} (\bibinfo{year}{2013}),
  \eprint{\href{http://arxiv.org/abs/1306.5546}{[arXiv:1306.5546 [hep-ph]]}}.

\bibitem[{\citenamefont{Davier et~al.}(2011)\citenamefont{Davier, Hoecker,
  Malaescu, and Zhang}}]{Davier2011}
\bibinfo{author}{\bibfnamefont{M.}~\bibnamefont{Davier}},
  \bibinfo{author}{\bibfnamefont{A.}~\bibnamefont{Hoecker}},
  \bibinfo{author}{\bibfnamefont{B.}~\bibnamefont{Malaescu}}, \bibnamefont{and}
  \bibinfo{author}{\bibfnamefont{Z.}~\bibnamefont{Zhang}},
  \bibinfo{journal}{Eur. Phys. J.} \textbf{\bibinfo{volume}{C71}},
  \bibinfo{pages}{1515} (\bibinfo{year}{2011}),
  \eprint{\href{http://arxiv.org/abs/1010.4180}{[arXiv:1010.4180 [hep-ph]]}}.

\bibitem[{\citenamefont{Hagiwara et~al.}(2011)\citenamefont{Hagiwara, Liao,
  Martin, Nomura, and Teubner}}]{Hagiwara2011}
\bibinfo{author}{\bibfnamefont{K.}~\bibnamefont{Hagiwara}},
  \bibinfo{author}{\bibfnamefont{R.}~\bibnamefont{Liao}},
  \bibinfo{author}{\bibfnamefont{A.~D.} \bibnamefont{Martin}},
  \bibinfo{author}{\bibfnamefont{D.}~\bibnamefont{Nomura}}, \bibnamefont{and}
  \bibinfo{author}{\bibfnamefont{T.}~\bibnamefont{Teubner}},
  \bibinfo{journal}{J. Phys.} \textbf{\bibinfo{volume}{G38}},
  \bibinfo{pages}{085003} (\bibinfo{year}{2011}),
  \eprint{\href{http://arxiv.org/abs/1105.3149}{[arXiv:1105.3149 [hep-ph]]}}.

\bibitem[{\citenamefont{Kurz et~al.}(2014)\citenamefont{Kurz, Liu, Marquard,
  and Steinhauser}}]{Kurz2014}
\bibinfo{author}{\bibfnamefont{A.}~\bibnamefont{Kurz}},
  \bibinfo{author}{\bibfnamefont{T.}~\bibnamefont{Liu}},
  \bibinfo{author}{\bibfnamefont{P.}~\bibnamefont{Marquard}}, \bibnamefont{and}
  \bibinfo{author}{\bibfnamefont{M.}~\bibnamefont{Steinhauser}},
  \bibinfo{journal}{Phys. Lett.} \textbf{\bibinfo{volume}{B734}},
  \bibinfo{pages}{144} (\bibinfo{year}{2014}),
  \eprint{\href{http://arxiv.org/abs/1403.6400}{[arXiv:1403.6400 [hep-ph]]}}.

\bibitem[{\citenamefont{Jegerlehner and Nyffeler}(2009)}]{Jegerlehner2009}
\bibinfo{author}{\bibfnamefont{F.}~\bibnamefont{Jegerlehner}} \bibnamefont{and}
  \bibinfo{author}{\bibfnamefont{A.}~\bibnamefont{Nyffeler}},
  \bibinfo{journal}{Phys. Rept.} \textbf{\bibinfo{volume}{477}},
  \bibinfo{pages}{1} (\bibinfo{year}{2009}),
  \eprint{\href{http://arxiv.org/abs/0902.3360}{[arXiv:0902.3360 [hep-ph]]}}.

\bibitem[{\citenamefont{Colangelo et~al.}(2014)\citenamefont{Colangelo,
  Hoferichter, Nyffeler, Passera, and Stoffer}}]{Colangelo2014}
\bibinfo{author}{\bibfnamefont{G.}~\bibnamefont{Colangelo}},
  \bibinfo{author}{\bibfnamefont{M.}~\bibnamefont{Hoferichter}},
  \bibinfo{author}{\bibfnamefont{A.}~\bibnamefont{Nyffeler}},
  \bibinfo{author}{\bibfnamefont{M.}~\bibnamefont{Passera}}, \bibnamefont{and}
  \bibinfo{author}{\bibfnamefont{P.}~\bibnamefont{Stoffer}},
  \bibinfo{journal}{Phys. Lett.} \textbf{\bibinfo{volume}{B735}},
  \bibinfo{pages}{90} (\bibinfo{year}{2014}),
  \eprint{\href{http://arxiv.org/abs/1403.7512}{[arXiv:1403.7512 [hep-ph]]}}.

\bibitem[{\citenamefont{St\"ockinger}(2007)}]{Stockinger:2006zn}
\bibinfo{author}{\bibfnamefont{D.}~\bibnamefont{St\"ockinger}},
  \bibinfo{journal}{J. Phys.} \textbf{\bibinfo{volume}{G34}},
  \bibinfo{pages}{R45} (\bibinfo{year}{2007}),
  \eprint{\href{http://arXiv.org/abs/hep-ph/0609168}{[arXiv:hep-ph/0609168]}}.

\bibitem[{\citenamefont{Chakraverty et~al.}(2001)\citenamefont{Chakraverty,
  Choudhury, and Datta}}]{Chakraverty:2001yg}
\bibinfo{author}{\bibfnamefont{D.}~\bibnamefont{Chakraverty}},
  \bibinfo{author}{\bibfnamefont{D.}~\bibnamefont{Choudhury}},
  \bibnamefont{and} \bibinfo{author}{\bibfnamefont{A.}~\bibnamefont{Datta}},
  \bibinfo{journal}{Phys. Lett.} \textbf{\bibinfo{volume}{B506}},
  \bibinfo{pages}{103} (\bibinfo{year}{2001}),
  \eprint{\href{http://arXiv.org/abs/hep-ph/0102180}{[arXiv:hep-ph/0102180]}}.

\bibitem[{\citenamefont{Cheung}(2001)}]{Cheung:2001ip}
\bibinfo{author}{\bibfnamefont{K.}~\bibnamefont{Cheung}},
  \bibinfo{journal}{Phys. Rev.} \textbf{\bibinfo{volume}{D64}},
  \bibinfo{pages}{033001} (\bibinfo{year}{2001}),
  \eprint{\href{http://arXiv.org/abs/hep-ph/0102238}{[arXiv:hep-ph/0102238]}}.

\bibitem[{\citenamefont{Freitas et~al.}(2014)\citenamefont{Freitas, Lykken,
  Kell, and Westhoff}}]{Freitas:2014pua}
\bibinfo{author}{\bibfnamefont{A.}~\bibnamefont{Freitas}},
  \bibinfo{author}{\bibfnamefont{J.}~\bibnamefont{Lykken}},
  \bibinfo{author}{\bibfnamefont{S.}~\bibnamefont{Kell}}, \bibnamefont{and}
  \bibinfo{author}{\bibfnamefont{S.}~\bibnamefont{Westhoff}},
  \bibinfo{journal}{JHEP} \textbf{\bibinfo{volume}{1405}}, \bibinfo{pages}{145}
  (\bibinfo{year}{2014}), \bibinfo{note}{[Erratum: JHEP {\bf 1409}, 155
  (2014)]}, \eprint{\href{http://arxiv.org/abs/1402.7065}{[arXiv:1402.7065
  [hep-ph]]}}.

\bibitem[{\citenamefont{Iltan and Sundu}(2003)}]{Iltan:2001nk}
\bibinfo{author}{\bibfnamefont{E.~O.} \bibnamefont{Iltan}} \bibnamefont{and}
  \bibinfo{author}{\bibfnamefont{H.}~\bibnamefont{Sundu}},
  \bibinfo{journal}{Acta Phys. Slov.} \textbf{\bibinfo{volume}{53}},
  \bibinfo{pages}{17} (\bibinfo{year}{2003}),
  \eprint{\href{http://arXiv.org/abs/hep-ph/0103105}{[arXiv:hep-ph/0103105]}}.

\bibitem[{\citenamefont{Broggio et~al.}(2014)\citenamefont{Broggio, Chun,
  Passera, Patel, and Vempati}}]{Broggio:2014mna}
\bibinfo{author}{\bibfnamefont{A.}~\bibnamefont{Broggio}},
  \bibinfo{author}{\bibfnamefont{E.~J.} \bibnamefont{Chun}},
  \bibinfo{author}{\bibfnamefont{M.}~\bibnamefont{Passera}},
  \bibinfo{author}{\bibfnamefont{K.~M.} \bibnamefont{Patel}}, \bibnamefont{and}
  \bibinfo{author}{\bibfnamefont{S.~K.} \bibnamefont{Vempati}},
  \bibinfo{journal}{JHEP} \textbf{\bibinfo{volume}{1411}}, \bibinfo{pages}{058}
  (\bibinfo{year}{2014}),
  \eprint{\href{http://arxiv.org/abs/1409.3199}{[arXiv:1409.3199 [hep-ph]]}}.

\bibitem[{\citenamefont{Wang and Han}(2015)}]{Wang:2014sda}
\bibinfo{author}{\bibfnamefont{L.}~\bibnamefont{Wang}} \bibnamefont{and}
  \bibinfo{author}{\bibfnamefont{X.-F.} \bibnamefont{Han}},
  \bibinfo{journal}{JHEP} \textbf{\bibinfo{volume}{1505}}, \bibinfo{pages}{039}
  (\bibinfo{year}{2015}),
  \eprint{\href{http://arxiv.org/abs/1412.4874}{[arXiv:1412.4874 [hep-ph]]}}.

\bibitem[{\citenamefont{Abe et~al.}(2015)\citenamefont{Abe, Sato, and
  Yagyu}}]{Abe:2015oca}
\bibinfo{author}{\bibfnamefont{T.}~\bibnamefont{Abe}},
  \bibinfo{author}{\bibfnamefont{R.}~\bibnamefont{Sato}}, \bibnamefont{and}
  \bibinfo{author}{\bibfnamefont{K.}~\bibnamefont{Yagyu}},
  \bibinfo{journal}{JHEP} \textbf{\bibinfo{volume}{1507}}, \bibinfo{pages}{064}
  (\bibinfo{year}{2015}),
  \eprint{\href{http://arxiv.org/abs/1504.07059}{[arXiv:1504.07059 [hep-ph]]}}.

\bibitem[{\citenamefont{Crivellin et~al.}(2016)\citenamefont{Crivellin, Heeck,
  and Stoffer}}]{Crivellin:2015hha}
\bibinfo{author}{\bibfnamefont{A.}~\bibnamefont{Crivellin}},
  \bibinfo{author}{\bibfnamefont{J.}~\bibnamefont{Heeck}}, \bibnamefont{and}
  \bibinfo{author}{\bibfnamefont{P.}~\bibnamefont{Stoffer}},
  \bibinfo{journal}{Phys. Rev. Lett.} \textbf{\bibinfo{volume}{116}},
  \bibinfo{pages}{081801} (\bibinfo{year}{2016}), \eprint{1507.07567}.

\bibitem[{\citenamefont{Langacker}(2009)}]{Langacker:2008yv}
\bibinfo{author}{\bibfnamefont{P.}~\bibnamefont{Langacker}},
  \bibinfo{journal}{Rev. Mod. Phys.} \textbf{\bibinfo{volume}{81}},
  \bibinfo{pages}{1199} (\bibinfo{year}{2009}),
  \eprint{\href{http://arxiv.org/abs/0801.1345}{[arXiv:0801.1345 [hep-ph]]}}.

\bibitem[{\citenamefont{Baek et~al.}(2001)\citenamefont{Baek, Deshpande, He,
  and Ko}}]{Baek:2001kca}
\bibinfo{author}{\bibfnamefont{S.}~\bibnamefont{Baek}},
  \bibinfo{author}{\bibfnamefont{N.~G.} \bibnamefont{Deshpande}},
  \bibinfo{author}{\bibfnamefont{X.~G.} \bibnamefont{He}}, \bibnamefont{and}
  \bibinfo{author}{\bibfnamefont{P.}~\bibnamefont{Ko}}, \bibinfo{journal}{Phys.
  Rev.} \textbf{\bibinfo{volume}{D64}}, \bibinfo{pages}{055006}
  (\bibinfo{year}{2001}),
  \eprint{\href{http://arXiv.org/abs/hep-ph/0104141}{[arXiv:hep-ph/0104141]}}.

\bibitem[{\citenamefont{Ma et~al.}(2002)\citenamefont{Ma, Roy, and
  Roy}}]{Ma:2001md}
\bibinfo{author}{\bibfnamefont{E.}~\bibnamefont{Ma}},
  \bibinfo{author}{\bibfnamefont{D.~P.} \bibnamefont{Roy}}, \bibnamefont{and}
  \bibinfo{author}{\bibfnamefont{S.}~\bibnamefont{Roy}},
  \bibinfo{journal}{Phys. Lett.} \textbf{\bibinfo{volume}{B525}},
  \bibinfo{pages}{101} (\bibinfo{year}{2002}),
  \eprint{\href{http://arXiv.org/abs/hep-ph/0110146}{[arXiv:hep-ph/0110146]}}.

\bibitem[{\citenamefont{Gninenko and Krasnikov}(2001)}]{Gninenko:2001hx}
\bibinfo{author}{\bibfnamefont{S.~N.} \bibnamefont{Gninenko}} \bibnamefont{and}
  \bibinfo{author}{\bibfnamefont{N.~V.} \bibnamefont{Krasnikov}},
  \bibinfo{journal}{Phys. Lett.} \textbf{\bibinfo{volume}{B513}},
  \bibinfo{pages}{119} (\bibinfo{year}{2001}),
  \eprint{\href{http://arXiv.org/abs/hep-ph/0102222}{[arXiv:hep-ph/0102222]}}.

\bibitem[{\citenamefont{Pospelov}(2009)}]{Pospelov:2008zw}
\bibinfo{author}{\bibfnamefont{M.}~\bibnamefont{Pospelov}},
  \bibinfo{journal}{Phys. Rev.} \textbf{\bibinfo{volume}{D80}},
  \bibinfo{pages}{095002} (\bibinfo{year}{2009}),
  \eprint{\href{http://arxiv.org/abs/0811.1030}{[arXiv:0811.1030 [hep-ph]]}}.

\bibitem[{\citenamefont{Heeck and Rodejohann}(2011)}]{Heeck:2011wj}
\bibinfo{author}{\bibfnamefont{J.}~\bibnamefont{Heeck}} \bibnamefont{and}
  \bibinfo{author}{\bibfnamefont{W.}~\bibnamefont{Rodejohann}},
  \bibinfo{journal}{Phys. Rev.} \textbf{\bibinfo{volume}{D84}},
  \bibinfo{pages}{075007} (\bibinfo{year}{2011}),
  \eprint{\href{http://arxiv.org/abs/1107.5238}{[arXiv:1107.5238 [hep-ph]]}}.

\bibitem[{\citenamefont{Harigaya et~al.}(2014)\citenamefont{Harigaya, Igari,
  Nojiri, Takeuchi, and Tobe}}]{Harigaya:2013twa}
\bibinfo{author}{\bibfnamefont{K.}~\bibnamefont{Harigaya}},
  \bibinfo{author}{\bibfnamefont{T.}~\bibnamefont{Igari}},
  \bibinfo{author}{\bibfnamefont{M.~M.} \bibnamefont{Nojiri}},
  \bibinfo{author}{\bibfnamefont{M.}~\bibnamefont{Takeuchi}}, \bibnamefont{and}
  \bibinfo{author}{\bibfnamefont{K.}~\bibnamefont{Tobe}},
  \bibinfo{journal}{JHEP} \textbf{\bibinfo{volume}{1403}}, \bibinfo{pages}{105}
  (\bibinfo{year}{2014}),
  \eprint{\href{http://arxiv.org/abs/1311.0870}{[arXiv:1311.0870 [hep-ph]]}}.

\bibitem[{\citenamefont{Altmannshofer
  et~al.}(2014{\natexlab{b}})\citenamefont{Altmannshofer, Gori, Pospelov, and
  Yavin}}]{Altmannshofer:2014pba}
\bibinfo{author}{\bibfnamefont{W.}~\bibnamefont{Altmannshofer}},
  \bibinfo{author}{\bibfnamefont{S.}~\bibnamefont{Gori}},
  \bibinfo{author}{\bibfnamefont{M.}~\bibnamefont{Pospelov}}, \bibnamefont{and}
  \bibinfo{author}{\bibfnamefont{I.}~\bibnamefont{Yavin}},
  \bibinfo{journal}{Phys. Rev. Lett.} \textbf{\bibinfo{volume}{113}},
  \bibinfo{pages}{091801} (\bibinfo{year}{2014}{\natexlab{b}}),
  \eprint{\href{http://arxiv.org/abs/1406.2332}{[arXiv:1406.2332 [hep-ph]]}}.

\bibitem[{\citenamefont{He et~al.}(1991{\natexlab{a}})\citenamefont{He, Joshi,
  Lew, and Volkas}}]{He:1990pn}
\bibinfo{author}{\bibfnamefont{X.~G.} \bibnamefont{He}},
  \bibinfo{author}{\bibfnamefont{G.~C.} \bibnamefont{Joshi}},
  \bibinfo{author}{\bibfnamefont{H.}~\bibnamefont{Lew}}, \bibnamefont{and}
  \bibinfo{author}{\bibfnamefont{R.~R.} \bibnamefont{Volkas}},
  \bibinfo{journal}{Phys. Rev.} \textbf{\bibinfo{volume}{D43}},
  \bibinfo{pages}{22} (\bibinfo{year}{1991}{\natexlab{a}}).

\bibitem[{\citenamefont{Foot}(1991)}]{Foot:1990mn}
\bibinfo{author}{\bibfnamefont{R.}~\bibnamefont{Foot}}, \bibinfo{journal}{Mod.
  Phys. Lett.} \textbf{\bibinfo{volume}{A6}}, \bibinfo{pages}{527}
  (\bibinfo{year}{1991}).

\bibitem[{\citenamefont{He et~al.}(1991{\natexlab{b}})\citenamefont{He, Joshi,
  Lew, and Volkas}}]{He:1991qd}
\bibinfo{author}{\bibfnamefont{X.-G.} \bibnamefont{He}},
  \bibinfo{author}{\bibfnamefont{G.~C.} \bibnamefont{Joshi}},
  \bibinfo{author}{\bibfnamefont{H.}~\bibnamefont{Lew}}, \bibnamefont{and}
  \bibinfo{author}{\bibfnamefont{R.~R.} \bibnamefont{Volkas}},
  \bibinfo{journal}{Phys. Rev.} \textbf{\bibinfo{volume}{D44}},
  \bibinfo{pages}{2118} (\bibinfo{year}{1991}{\natexlab{b}}).

\bibitem[{\citenamefont{Binetruy et~al.}(1997)\citenamefont{Binetruy, Lavignac,
  Petcov, and Ramond}}]{Binetruy:1996cs}
\bibinfo{author}{\bibfnamefont{P.}~\bibnamefont{Binetruy}},
  \bibinfo{author}{\bibfnamefont{S.}~\bibnamefont{Lavignac}},
  \bibinfo{author}{\bibfnamefont{S.~T.} \bibnamefont{Petcov}},
  \bibnamefont{and} \bibinfo{author}{\bibfnamefont{P.}~\bibnamefont{Ramond}},
  \bibinfo{journal}{Nucl. Phys.} \textbf{\bibinfo{volume}{B496}},
  \bibinfo{pages}{3} (\bibinfo{year}{1997}), \eprint{hep-ph/9610481}.

\bibitem[{\citenamefont{Bell and Volkas}(2001)}]{Bell:2000vh}
\bibinfo{author}{\bibfnamefont{N.~F.} \bibnamefont{Bell}} \bibnamefont{and}
  \bibinfo{author}{\bibfnamefont{R.~R.} \bibnamefont{Volkas}},
  \bibinfo{journal}{Phys. Rev.} \textbf{\bibinfo{volume}{D63}},
  \bibinfo{pages}{013006} (\bibinfo{year}{2001}), \eprint{hep-ph/0008177}.

\bibitem[{\citenamefont{Choubey and Rodejohann}(2005)}]{Choubey:2004hn}
\bibinfo{author}{\bibfnamefont{S.}~\bibnamefont{Choubey}} \bibnamefont{and}
  \bibinfo{author}{\bibfnamefont{W.}~\bibnamefont{Rodejohann}},
  \bibinfo{journal}{Eur. Phys. J.} \textbf{\bibinfo{volume}{C40}},
  \bibinfo{pages}{259} (\bibinfo{year}{2005}), \eprint{hep-ph/0411190}.

\bibitem[{\citenamefont{Dutta et~al.}(1994)\citenamefont{Dutta, Joshipura, and
  Vijaykumar}}]{Dutta:1994dx}
\bibinfo{author}{\bibfnamefont{G.}~\bibnamefont{Dutta}},
  \bibinfo{author}{\bibfnamefont{A.~S.} \bibnamefont{Joshipura}},
  \bibnamefont{and} \bibinfo{author}{\bibfnamefont{K.~B.}
  \bibnamefont{Vijaykumar}}, \bibinfo{journal}{Phys. Rev.}
  \textbf{\bibinfo{volume}{D50}}, \bibinfo{pages}{2109} (\bibinfo{year}{1994}),
  \eprint{hep-ph/9405292}.

\bibitem[{con()}]{confnote}
\bibinfo{note}{{ATLAS and CMS Collaboration, ATLAS-CONF-2015-044}}.

\bibitem[{\citenamefont{Aad et~al.}(2016{\natexlab{b}})}]{Aad:2015gba}
\bibinfo{author}{\bibfnamefont{G.}~\bibnamefont{Aad}} \bibnamefont{et~al.}
  (\bibinfo{collaboration}{ATLAS}), \bibinfo{journal}{Eur. Phys. J.}
  \textbf{\bibinfo{volume}{C76}}, \bibinfo{pages}{6}
  (\bibinfo{year}{2016}{\natexlab{b}}), \eprint{1507.04548}.

\bibitem[{\citenamefont{Khachatryan
  et~al.}(2015{\natexlab{b}})}]{Khachatryan:2014jba}
\bibinfo{author}{\bibfnamefont{V.}~\bibnamefont{Khachatryan}}
  \bibnamefont{et~al.} (\bibinfo{collaboration}{CMS}), \bibinfo{journal}{Eur.
  Phys. J.} \textbf{\bibinfo{volume}{C75}}, \bibinfo{pages}{212}
  (\bibinfo{year}{2015}{\natexlab{b}}), \eprint{1412.8662}.

\bibitem[{\citenamefont{Robens and Stefaniak}(2015)}]{Robens:2015gla}
\bibinfo{author}{\bibfnamefont{T.}~\bibnamefont{Robens}} \bibnamefont{and}
  \bibinfo{author}{\bibfnamefont{T.}~\bibnamefont{Stefaniak}},
  \bibinfo{journal}{Eur. Phys. J.} \textbf{\bibinfo{volume}{C75}},
  \bibinfo{pages}{104} (\bibinfo{year}{2015}), \eprint{1501.02234}.

\bibitem[{\citenamefont{Khachatryan
  et~al.}(2015{\natexlab{c}})}]{Khachatryan:2015cwa}
\bibinfo{author}{\bibfnamefont{V.}~\bibnamefont{Khachatryan}}
  \bibnamefont{et~al.} (\bibinfo{collaboration}{CMS}), \bibinfo{journal}{JHEP}
  \textbf{\bibinfo{volume}{10}}, \bibinfo{pages}{144}
  (\bibinfo{year}{2015}{\natexlab{c}}), \eprint{1504.00936}.

\bibitem[{\citenamefont{Aad et~al.}(2016{\natexlab{c}})}]{Aad:2015kna}
\bibinfo{author}{\bibfnamefont{G.}~\bibnamefont{Aad}} \bibnamefont{et~al.}
  (\bibinfo{collaboration}{ATLAS}), \bibinfo{journal}{Eur. Phys. J.}
  \textbf{\bibinfo{volume}{C76}}, \bibinfo{pages}{45}
  (\bibinfo{year}{2016}{\natexlab{c}}), \eprint{1507.05930}.

\bibitem[{\citenamefont{Andersen et~al.}(2013)}]{Heinemeyer:2013tqa}
\bibinfo{author}{\bibfnamefont{J.~R.} \bibnamefont{Andersen}}
  \bibnamefont{et~al.} (\bibinfo{collaboration}{LHC Higgs Cross Section Working
  Group}) (\bibinfo{year}{2013}), \eprint{1307.1347}.

\bibitem[{\citenamefont{Khachatryan
  et~al.}(2015{\natexlab{d}})}]{Khachatryan:2014aep}
\bibinfo{author}{\bibfnamefont{V.}~\bibnamefont{Khachatryan}}
  \bibnamefont{et~al.} (\bibinfo{collaboration}{CMS}), \bibinfo{journal}{Phys.
  Lett.} \textbf{\bibinfo{volume}{B744}}, \bibinfo{pages}{184}
  (\bibinfo{year}{2015}{\natexlab{d}}), \eprint{1410.6679}.

\bibitem[{\citenamefont{Aubert et~al.}(2010)}]{Aubert:2009ag}
\bibinfo{author}{\bibfnamefont{B.}~\bibnamefont{Aubert}} \bibnamefont{et~al.}
  (\bibinfo{collaboration}{BaBar}), \bibinfo{journal}{Phys. Rev. Lett.}
  \textbf{\bibinfo{volume}{104}}, \bibinfo{pages}{021802}
  (\bibinfo{year}{2010}),
  \eprint{\href{http://arxiv.org/abs/0908.2381}{[arXiv:0908.2381 [hep-ex]]}}.

\bibitem[{\citenamefont{Hayasaka et~al.}(2008)}]{Hayasaka:2007vc}
\bibinfo{author}{\bibfnamefont{K.}~\bibnamefont{Hayasaka}} \bibnamefont{et~al.}
  (\bibinfo{collaboration}{Belle}), \bibinfo{journal}{Phys. Lett.}
  \textbf{\bibinfo{volume}{B666}}, \bibinfo{pages}{16} (\bibinfo{year}{2008}),
  \eprint{\href{http://arxiv.org/abs/0705.0650}{[arXiv:0705.0650 [hep-ex]]}}.

\bibitem[{\citenamefont{Aushev et~al.}(2010)}]{Aushev:2010bq}
\bibinfo{author}{\bibfnamefont{T.}~\bibnamefont{Aushev}} \bibnamefont{et~al.}
  (\bibinfo{year}{2010}), \eprint{1002.5012}.

\bibitem[{\citenamefont{Bauer et~al.}(2015{\natexlab{a}})\citenamefont{Bauer,
  Carena, and Gemmler}}]{Bauer:2015fxa}
\bibinfo{author}{\bibfnamefont{M.}~\bibnamefont{Bauer}},
  \bibinfo{author}{\bibfnamefont{M.}~\bibnamefont{Carena}}, \bibnamefont{and}
  \bibinfo{author}{\bibfnamefont{K.}~\bibnamefont{Gemmler}},
  \bibinfo{journal}{JHEP} \textbf{\bibinfo{volume}{11}}, \bibinfo{pages}{016}
  (\bibinfo{year}{2015}{\natexlab{a}}), \eprint{1506.01719}.

\bibitem[{\citenamefont{Bauer et~al.}(2015{\natexlab{b}})\citenamefont{Bauer,
  Carena, and Gemmler}}]{Bauer:2015kzy}
\bibinfo{author}{\bibfnamefont{M.}~\bibnamefont{Bauer}},
  \bibinfo{author}{\bibfnamefont{M.}~\bibnamefont{Carena}}, \bibnamefont{and}
  \bibinfo{author}{\bibfnamefont{K.}~\bibnamefont{Gemmler}}
  (\bibinfo{year}{2015}{\natexlab{b}}), \eprint{1512.03458}.

\bibitem[{\citenamefont{Huitu et~al.}(2016)\citenamefont{Huitu, Keus, Koivunen,
  and Lebedev}}]{Huitu:2016pwk}
\bibinfo{author}{\bibfnamefont{K.}~\bibnamefont{Huitu}},
  \bibinfo{author}{\bibfnamefont{V.}~\bibnamefont{Keus}},
  \bibinfo{author}{\bibfnamefont{N.}~\bibnamefont{Koivunen}}, \bibnamefont{and}
  \bibinfo{author}{\bibfnamefont{O.}~\bibnamefont{Lebedev}}
  (\bibinfo{year}{2016}), \eprint{1603.06614}.

\bibitem[{dip({\natexlab{a}})}]{diphoton1}
\bibinfo{note}{{ATLAS Collaboration, ATLAS-CONF-2015-081}}.

\bibitem[{dip({\natexlab{b}})}]{diphoton2}
\bibinfo{note}{{CMS Collaboration, CMS-PAS-EXO-15-004}}.

\end{thebibliography}

\end{document}